\documentclass[%
reprint,
superscriptaddress,
showpacs,preprintnumbers,showkeys,
amsmath,amssymb,
aps,
prb,
floatfix,
]{revtex4-1}

\usepackage{graphicx}% Include figure files
\usepackage{dcolumn}% Align table columns on decimal point
\usepackage{bm}% bold math
\usepackage{hyperref}% add hypertext capabilities
\usepackage[mathlines]{lineno}% Enable numbering of text and display math

\usepackage{multirow}
\usepackage[capitalize]{cleveref} % Allows reference of multiple equations at once
\usepackage{placeins}
\usepackage{xcolor}

\newcommand{\br}{\mathbf{r}}
\newcommand{\bra}[1]{\langle #1|}

\newcommand{\ket}[1]{|#1\rangle}
\newcommand{\nline}{\nonumber \\}
\newcommand{\Trace}[1]{\mathrm{Tr}\left[#1\right]}

\newcommand{\edit}[1]{#1}

\begin{document}

%\preprint{APS/123-QED}

\title{The role of spin in the calculation of Hubbard $U$ and Hund's $J$ parameters from first principles}

\author{Edward B.\ Linscott}
\email{ebl27@cam.ac.uk}
\affiliation{Cavendish Laboratory, University of Cambridge, J.\ J.\ Thomson Avenue, Cambridge CB3 0HE, United Kingdom}
\author{Daniel J.\ Cole}
\affiliation{School of Natural and Environmental Sciences, Newcastle University, Newcastle upon Tyne NE1 7RU, United Kingdom}
\author{Michael C.\ Payne}
\affiliation{Cavendish Laboratory, University of Cambridge, J.\ J.\ Thomson Avenue, Cambridge CB3 0HE, United Kingdom}
\author{David D.\ O'Regan}
\email{david.o.regan@tcd.ie}
\affiliation{School of Physics, CRANN and AMBER, Trinity College Dublin, Dublin 2, Ireland}

\date{\today}

\begin{abstract}
The density functional theory (DFT)\,+\,$U$ method is a pragmatic and effective approach for calculating the ground-state properties of strongly-correlated systems, and linear response calculations are widely used to determine the requisite Hubbard parameters from first principles. We provide a detailed treatment of spin within this linear response approach, \edit{demonstrating that the conventional Hubbard $U$ formula, unlike the conventional DFT\,+\,$U$ corrective functional, incorporates interactions that are off-diagonal in the spin indices and places greater weight on one spin channel over the other.} We construct alternative definitions for Hubbard and Hund's parameters that are consistent with the contemporary DFT\,+\,$U$ functional, expanding upon the minimum-tracking linear response method. \edit{This approach allows Hund's $J$ and spin-dependent $U$ parameters to be calculated with the same ease as for the standard Hubbard $U$. Our methods accurately reproduce the experimental band gap, local magnetic moments, and the valence band edge character of manganese oxide, a canonical strongly-correlated system. We also} apply our approach to a complete series of transition-metal complexes [M(H\textsubscript{2}O)\textsubscript{6}]\textsuperscript{n+} (for M = Ti to Zn), showing that Hubbard corrections on oxygen atoms are necessary for preserving bond lengths, and demonstrating that our methods are numerically well-behaved even for near-filled subspaces such as in zinc. However, spectroscopic properties appear beyond the reach of the standard DFT\,+\,$U$ approach. \edit{Collectively, these results shed new light on the role of spin in the calculation of the corrective parameters $U$ and $J$, and point the way towards avenues for further development of DFT\,+\,$U$-type methods.}
\end{abstract}

\pacs{31.15.-p, 31.15.ej, 31.15.es}
% See https://publishing.aip.org/publishing/pacs/pacs-reg30#31
% \keywords{density functional theory, DFT+U, linear response}
%Use showkeys class option if keyword
\maketitle

% \tableofcontents

%%%%%%%%%%%%%%%%%%%%%%%%%%%%%%%%%%%%%%%%%%%%%%%%%%%%%%%%%%%%%%%%%%%%%%%%%%%%%%%
\section{\label{sec:intro}Introduction}
%%%%%%%%%%%%%%%%%%%%%%%%%%%%%%%%%%%%%%%%%%%%%%%%%%%%%%%%%%%%%%%%%%%%%%%%%%%%%%%
Over the past few decades, density functional theory (DFT) has played a key role in the simulation of many-body atomistic systems.\cite{Jones2015a, Jain2016a} DFT makes such systems tractable via the Hohenberg-Kohn theorems\cite{Hohenberg1964a} and the Kohn-Sham construction,\cite{Kohn1965a} but exchange and correlation must be approximated in the form of an exchange-correlation (xc)-functional.\cite{Perdew1981a, Perdew1996a}% Common xc-functionals such as the local-density approximation (LDA) \cite{Perdew1981a} and generalised gradient approximations (GGAs) \cite{Perdew1996a} are adequate for many systems, but become unreliable when treating systems that exhibit significant electronic correlation.

% One approach for improving upon these entry-level xc-functionals is Hubbard-augmented DFT (or DFT\,+\,$U$) \cite{Dudarev1998a, Anisimov1991b, Anisimov1991a, Anisimov1993a, Pickett1998a, Anisimov1997a}. Inspired by the success of the seminal Hubbard model \cite{Hubbard1963a}, DFT\,+\,$U$ involves adding Hubbard-model-like terms to the DFT framework. The added terms are applied to spatially localized subsystems that are expected to exhibit strong correlation  --- for instance, the $3d$ orbitals of a transition metal --- while the rest of the system is treated at the standard DFT level.
% 
% Historically, the DFT\,+\,$U$ formalism was developed by adapting the Hubbard model in order for it to be incorporated into the framework of DFT --- in the hope that existing xc-functionals would inherit the Hubbard model's ability to describe the Mott transition and other related physics. However, a number of steps in its derivation are hard to rigorously justify, and consequently this interpretation of DFT\,+\,$U$ has fallen out of favor. Nevertheless, the technique itself remains popular thanks to a reinterpretation of the Hubbard correction thanks to the work of Cococcioni, Kulik, and co-workers \cite{Cococcioni2005a, Kulik2006a}.
%
%The original goal that DFT\,+\,$U$ set out to achieve was to improve DFT's treatment of exchange and correlation.
One of the most prominent failures of many xc-functionals is that they do not properly correct for the self-interaction in the Hartree term. Self-interaction error (SIE) --- or more generally ``delocalization error''\cite{Cohen2008a,Cohen2012a} --- manifests itself as a spurious curvature in total energies with respect to total electron number, where instead there should be a derivative discontinuity at integer numbers of electrons and linear behavior at fractional numbers.\cite{Perdew1982a} This failure is closely related to approximate DFT's well-documented underestimation of the band gap.\cite{Cococcioni2005a, Kaduk2012a,Mori-Sanchez2008a}

While the origins of the SIE are well understood, it remains a challenge to avoid its introduction when constructing xc-functionals, even if exact exchange is incorporated.\cite{Cohen2008a} A simple yet remarkably successful alternative is Hubbard-augmented DFT (LDA\,+\,$U$ or more generally DFT\,+\,$U$). In this scheme, Hubbard model terms are incorporated into the DFT framework. This approach was originally designed to capture Mott-Hubbard physics in transition-metal oxides,\cite{Anisimov1991a, Anisimov1991a, Dudarev1998a} but it has subsequently gained a transparent interpretation as a corrective method for SIE due to the work of Cococcioni, Kulik, and co-workers.\cite{Cococcioni2005a, Kulik2006a} \edit{They observed that Hubbard corrections can counteract the spurious SIE curvature --- in other words, DFT\,+\,$U$ calculations may be constructed to cancel the SIE that is present (although this is not guaranteed).\cite{Zhao2016a}}

\subsection{The DFT\,+\,U correction}

In the DFT\,+\,$U$ scheme, one adds to the energy a corrective term (here we use the rotationally-invariant, simplified form),\cite{Dudarev1998a, Anisimov1991b, Anisimov1991a, Anisimov1993a, Anisimov1997a, Pickett1998a} given by
\begin{align}
E_\text{U}[\hat n^{I\sigma}] = \sum_{I\sigma} \frac{U^I}{2}\Trace{\hat n^{I\sigma}(1-\hat n^{I\sigma})},
\label{eqn:DFT+U_energy}
\end{align}
where the density operators $\hat n^{I\sigma} = \hat P^I \hat \rho^{\sigma} \hat P^I$ are projections of the (spin-dependent) Kohn-Sham density operator onto subspaces (indexed $I$) in which the SIE is to be addressed. The projectors ${\hat P^I=\sum_m \ket{\varphi^I_m}\bra{\varphi^{Im}}}$ are typically constructed from atom-centered, fixed, spin-independent, localized, and orthonormal orbitals $\varphi^I_m$ (although they may be non-orthogonal\cite{ORegan2011a} and self-consistent\cite{ORegan2010a}). 
% Ideally, the resulting subspaces should be relatively non-interacting with their environment: intra-subspace interactions should be much larger than any others if corrections to curvatures with respect to subspace (\emph{c.f.}\ total) electron number are going to address curvatures with respect to total electron number.
The $U^I$ are externally-defined parameters that determine the strength of the energy corrections. If they are well-chosen, the term that is quadratic in $\hat n$ can \edit{partially} correct the spurious energy curvature arising from the SIE.\cite{Zhao2016a} In the basis of localized orbitals $\psi_m^{I\sigma}$ that diagonalize the \edit{subspace occupancy} matrices such that $\hat n^{I\sigma} \psi_m^{I\sigma} = \lambda_m^{I\sigma} \psi_m^{I\sigma}$, the Hubbard correction becomes $\sum_{I\sigma m} U^I\lambda_m^{I\sigma}(1- \lambda_m^{I\sigma})/2$, which penalizes non-integer occupancies of these orbitals $\psi_m^{I\sigma}$. The explicit correction to the total energy vanishes at integer occupancy \edit{matrix eigenvalues}, where the xc-functional is assumed to be correct.

The corresponding correction to the Kohn-Sham potential is given by
\begin{align}
\hat V_\text{U} = \sum_{I\sigma m n}U^{I} \ket{\varphi^I_m}\left(\frac{1}{2}-{n^{I\sigma m}}_n \right) \bra{\varphi^{In}}.
\label{eqn:DFT+U_potential}
\end{align}
This is attractive or repulsive for occupancy matrix eigenvalues greater than or less than one-half, respectively. In the absence of any significant self-consistent response, this will penalize non-integer occupancies of the subspaces, opening an energy gap of order $U$ between any occupied and unoccupied Kohn-Sham orbitals with significant overlap with the Hubbard projectors. 

In order to correct interactions between unlike spins, DFT\,+\,$U$ can be extended to \edit{become} DFT\,+\,$U$\,+\,$J$ .\cite{Solovyev1994a, Mosey2007a, Himmetoglu2011a, Himmetoglu2014a} This involves a second correction to the total energy,
\begin{align}
E_\text{J}[\hat n^\sigma] = \sum_{I\sigma}\frac{J^I}{2}\Trace{\hat n^{I\sigma}\hat n^{I-\sigma}},
\label{eqn:DFT+J_energy}
\end{align}
where this correction is parameterized by the additional \edit{Hund's coupling} constants $J^I$. Additionally, the $U$ in Eqs.~\ref{eqn:DFT+U_energy} and \ref{eqn:DFT+U_potential} becomes $U_\mathrm{eff} = U - J$.

A substantial advantage of DFT\,+\,$U$\,(+\,$J$) over other methods that address the SIE (for example, SIC-LSDA,\cite{Lueders2005a, Dane2009a, Hughes2007a} Fermi orbital self-interaction correction,\cite{Pederson2014a, Pederson2015a, Hahn2015a} and Koopman's compliant functionals\cite{Dabo2010a, Borghi2014a}) is its small computational cost: once any Hubbard parameters have been determined, the overhead for incorporating the additional potential and energy terms is insignificant compared to the cost of the DFT calculation itself.\cite{ORegan2012a}

%%%%%%%%%%%%%%%%%%%%%%%%%%%%%%%%%%%%%%%%%%%%%%%%%%%%%%%%%%%%%%%%%%%%%%%%%%%%%%%
\subsection{\label{subsec:Cococcioni_approach} Conventional linear response}
%%%%%%%%%%%%%%%%%%%%%%%%%%%%%%%%%%%%%%%%%%%%%%%%%%%%%%%%%%%%%%%%%%%%%%%%%%%%%%%
In order to apply a Hubbard correction, one must select an appropriate value for the parameters $U^I$. This can be done pragmatically by picking values on empirical grounds --- that is, chosen so that certain system characteristics are reproduced (for example, ionic geometries,\cite{Loschen2007a, Castleton2007a, Morgan2007a, Franchini2007a} band gaps,\cite{Castleton2007a, Morgan2007a, Rohrbach2004a, Franchini2007a} and formation enthalpies\cite{Franchini2007a, Wang2006a, Ong2011a}). While this approach has seen some success,\cite{Fennie2006a,Finazzi2008a} it does not guarantee that the chosen $U$ will correct the SIE energy curvature \edit{to the greatest extent achievable}, or result in \edit{an} improved description of other system properties, and is not even possible where there is a lack of reliable experimental or higher-level computational data.

An alternative approach is the linear response method developed by Cococcioni and de Gironcoli,\cite{Cococcioni2005a} which built upon the earlier linear response scheme of Pickett and co-workers,\cite{Pickett1998a} and shares many aspects with the constrained LDA approach of Aryasetiawan and co-workers.\cite{Aryasetiawan2006a} In this approach, DFT calculations are performed subject to a perturbing potential $\delta \hat v_\mathrm{ext} = dv^J_\mathrm{ext} \hat P^J$ confined to the $J$\textsuperscript{th} Hubbard subspace, for a range of values of $dv^J_\mathrm{ext}$. The density operator's response to these perturbations is given by the response operator $\hat \chi$:
\begin{equation}
   \delta \hat \rho = \hat \chi \delta \hat v_\mathrm{ext}.
\end{equation}
The occupancy of the $I$\textsuperscript{th} Hubbard subspace will change by
\begin{equation}
   dn^I = \Trace{\hat P^I \delta \hat \rho}
        = \Trace{\hat P^I \hat \chi \hat P^J} dv^J_\mathrm{ext}
\end{equation}
and thus we can define the projected response matrix\cite{ORegan2016a}
\begin{equation}
   \chi_{IJ} \equiv \frac{dn^I}{dv^J_\mathrm{ext}} = \Trace{\hat P^I \hat \chi \hat P^J}.
\end{equation}
% \begin{equation}
% \chi_{IJ} = \frac{\partial n^I}{\partial \alpha^J},
% %\chi_{IJ} = \frac{\partial ^2\mathcal{E}[\{\alpha_K\}]}{\partial \alpha_J \partial \alpha_I} = \frac{\partial}{\partial \alpha_J}\frac{\partial \mathcal{E}[\{\alpha_K\}]}{ \partial \alpha_I} = \frac{\partial n_I}{\partial \alpha_J}.
% \label{eqn:U_linear_response_chiIJ}
% \end{equation}
%
% where $n^I = \sum_\sigma \Trace{\hat n^{I\sigma}}$.
%
% \begin{align}
% \mathcal{E}[\{\alpha_I\}] = \min_{n(\br)}\left\{E[n(\br)]+\sum_I \alpha_I n_I \right\}.
% \end{align}
%
A value for $U$ that corresponds to the screened response of the system is given by
\begin{equation}
U^I =  \left(\chi^{-1}_0 - \chi^{-1} \right)_{II}
\label{eqn:U_linear_response_UminusU0}
\end{equation}
where $\chi_0$ is the response of the non-interacting system, which must be separately measured and removed from the Hubbard correction.\cite{CococcioniThesis,Dederichs1984a,Himmetoglu2014a} There is also scope here for calculation of off-diagonal terms $V_{IJ} = \left(\chi^{-1}_0 - \chi^{-1} \right)_{IJ}$, which gives rise to DFT\,+\,$U$\,+\,$V$.\cite{Campo2010a, Kulik2011a} %This non-interacting is included because transpires that if one perturbed a non-interacting system, we would still obtain quadratic behaviour \cite{Cococcioni2005a}. Because this curvature is not associated with electron-electron interactions, correcting for it via a $U$-like term is not considered to be sensible. Consequently, the non-interacting response must be measured and removed from the Hubbard correction.

Satisfyingly, the determination of $U$ via linear response removes any possible arbitrariness of the Hubbard correction: the $U$ parameter is a well-defined property of the system that can be unambiguously measured in theory.\cite{Kulik2006a, Himmetoglu2014a}

\edit{Recently the idea of calculating $U$ and $J$ to best emulate subspace-projected Kohn-Sham exact exchange\cite{Mosey2007a} has been further advanced.\cite{Agapito2015a} But because we wish to cancel the systematic errors of approximate DFT\cite{Perdew1982a,Cohen2012a} to the extent possible using functionals of the DFT\,+\,$U$ form, we choose to instead develop the linear-response formalism that has been shown to successfully achieve this,\cite{Pickett1998a,Cococcioni2005a,Kulik2006a,Moynihan2017a} and that does not incorporate any theory or model (\emph{e.g.} Fock exchange) beyond what is already ordinarily present.}

%%%%%%%%%%%%%%%%%%%%%%%%%%%%%%%%%%%%%%%%%%%%%%%%%%%%%%%%%%%%%%%%%%%%%%%%%%%%%%%
\subsection{Problem and paper outline}
%%%%%%%%%%%%%%%%%%%%%%%%%%%%%%%%%%%%%%%%%%%%%%%%%%%%%%%%%%%%%%%%%%%%%%%%%%%%%%%
There are some aspects of the linear response methodology that pose issues. Firstly, delocalization error is associated with fractional total charge, but the DFT\,+\,$U$ functional of Eq.~\ref{eqn:DFT+U_energy} corrects fractional occupation for each spin channel separately. Conventional linear response, meanwhile, perturbs both spin channels simultaneously. These discrepancies in how we treat spin channels warrant investigation.

Secondly, measuring the non-interacting response $\chi_0$ is not straightforward. The common practice is to follow the example of Ref.~\onlinecite{Cococcioni2005a}, and calculate $\chi_0$ via the first iteration of the Kohn-Sham equations during a self-consistent field calculation --- that is, the response is to be measured following the initial charge redistribution introduced by the perturbation but before the Kohn-Sham potential is updated. This approach is impractical to implement in codes that use a direct-minimization procedure of the total energy with respect to the density, Kohn-Sham orbitals, or density-matrix. This represents a substantial number of packages, including ONETEP,\cite{Skylaris2005a} CONQUEST,\cite{Bowler2002a, Gillan2007a} SIESTA,\cite{Soler2002a, Artacho2008a} \textsc{BigDFT},\cite{Genovese2011a} \textsc{OpenMX},\cite{Ozaki2005a} and CP2K\cite{Weber2008a} \edit{(albeit that in some of these the self-consistent field technique is also available).} In direct-minimization, updating the density and potential are not nested separately, so $\chi_0$ cannot be calculated in the manner prescribed above.%\textcolor{red}{Could also mention that defined in this way $U$ is not strictly a ground-state property, or could be a bit more sensitive and wait till later to include this point}

Finally, in the case of closed-shell systems, linear response approaches tend to dramatically overestimate $U$ and suffer from numerical instabilities.\cite{Hu2006a,Kulik2010a,Yu2014a,Huang2017a} We would like to overcome, or at least to better understand, these failures.

In Ref.~\onlinecite{Moynihan2017a} one of us presented an alternative to the established SCF linear response approach for calculating the Hubbard parameters. This minimum-tracking linear response approach is suited for calculating $U$ in direct-minimization codes. In section~\ref{sec:theory}, we expand upon this formalism. To resolve the discrepancies between conventional linear response and the contemporary DFT\,+\,$U$ functional, we pay particular attention to spin and associated screening, proposing revised definitions for Hubbard and Hund's parameters (section~\ref{subsec:spin_complications}). By comparing scalar linear response to our spin-specific theory, we demonstrate that the treatment of inter-spin screening in conventional linear response is somewhat inconsistent with the DFT\,+\,$U$ functional as it is most commonly employed (section~\ref{subsec:comparisons_with_conventional_approach}). \edit{While we do not claim here to arrive at an ultimate solution to this inconsistency, we do provide a simple technique by which inter-spin screening of the Hubbard $U$ may be suppressed. This results in spin-dependent $U$ parameters that are generally lower in value than the canonical $U$ for the partially-filled spin channel of a localized subspace (the spin channel that usually harbours the strong correlation effects) and that, in principle, could be applied to that spin channel alone. This hints at a possible solution to the widespread finding that first-principles $U$ parameters can be rather too large, in practice, leading to over-correction by DFT\,+\,$U$.}

In the latter half of the paper (section~\ref{sec:application}) we apply our theoretical developments to a complete set of hexahydrated transition metal complexes from Ti to Zn. We calculate Hubbard and Hund's parameters using conventional and novel approaches (section~\ref{subsec:hexX_Hubbard_parameters}), and then perform DFT\,+\,$U$ calculations using these parameters to predict structural and spectroscopic properties (\cref{subsec:hexX_structural,subsec:hexX_spectroscopic}). The numerical stability of the minimum-tracking formalism (in which Hubbard parameters are a strictly ground state property) allows us to investigate closed-shell cases with confidence. The Hubbard corrections to oxygen $2p$ subspaces are far from negligible, \edit{and help to obtain sensible structural predictions. Spectroscopic simulations of coordination complexes using DFT\,+\,$U$ see only mixed success, whereas our indirect band gap results for the long-standing challenge material MnO are very promising when compared against a wide range of more computationally demanding approximations.}

%%%%%%%%%%%%%%%%%%%%%%%%%%%%%%%%%%%%%%%%%%%%%%%%%%%%%%%%%%%%%%%%%%%%%%%%%%%%%%%
\section{\label{sec:theory} The minimum-tracking approach for calculating U via linear response}
%%%%%%%%%%%%%%%%%%%%%%%%%%%%%%%%%%%%%%%%%%%%%%%%%%%%%%%%%%%%%%%%%%%%%%%%%%%%%%%
The minimum-tracking linear response approach is largely equivalent to SCF linear response, but its derivation centers on the ground-state density for each value of the perturbing potential.

As with the SCF approach, a perturbing potential $d \hat v_\mathrm{ext} = dv^J_\mathrm{ext}\hat P^J$ is applied to the $J$\textsuperscript{th} Hubbard subspace. The response of the projected Kohn-Sham potential is given by the chain rule
% The kernel, including off diagonal terms, is given by
% %, where the projection of a one-body operator $\hat O$ onto the $I$\textsuperscript{th} Hubbard subspace is defined as 
% %
% \begin{equation}
%    f_{IJ} = \frac{dv^I_\mathrm{Hxc}}{dn^J} = \frac{dv^I_\mathrm{KS}}{dn^J} - \frac{dv^I_\mathrm{ext}}{dn^J}.
%    \label{eqn:variational_Dyson_equation}
% \end{equation}
% %
% This is nothing more than a reiteration of eq.~\ref{eqn:U_linear_response_UminusU0}; the two terms can be recognised as $(\chi^{-1}_0)_{IJ}$ and $(\chi^{-1})_{IJ}$, as consistent with standard linear response theory (see Appendix \ref{sec:chi0_in_detail}). The removal of the non-interacting response can be rigorously justified as a consequence of the Dyson equation, with $U$ being a measure of net interaction.
%
%Changes in the Hartree-plus-xc potential are related to changes in the Kohn-Sham potential via $dv_\mathrm{Hxc} = dv_\mathrm{KS} - dv_\mathrm{ext}$. In a linear response calculation, the response of the projected Kohn-Sham potential is given by the chain rule
%
\begin{equation}
   \frac{dv^I_\mathrm{KS} }{dv^J_\mathrm{ext}}
 = \frac{dv^I_\mathrm{ext}}{dv^J_\mathrm{ext}}
 + \frac{dv^I_\mathrm{Hxc}}{dv^J_\mathrm{ext}}
 = \frac{dv^I_\mathrm{ext}}{dv^J_\mathrm{ext}}
 + \sum_K\frac{dv^I_\mathrm{Hxc}}{dn^K}\frac{dn^K}{dv^J_\mathrm{ext}},
\label{eqn:Hubbard_chain_rule}
\end{equation}
where the final step follows because while the external potential acting on site $J$ will change the density matrix everywhere, the $N$-site Hubbard model only sees the $N$ subspace density matrices. Screening due to the residual bath is incorporated within the total derivatives. The projections of one-body operators are given by $O^I = \mathrm{Tr}[\hat P^I \hat O]/\mathrm{Tr}[\hat P^I].$

Defining 
$f_{IJ} \equiv d v^I_\mathrm{Hxc}/d n^J$, $(\varepsilon^{-1})_{IJ} \equiv d v^I_\mathrm{KS}/d v^J_\mathrm{ext}$, and $\Omega_{IJ} \equiv d v^I_\mathrm{ext}/d v^J_\mathrm{ext}$, Eq.~\ref{eqn:Hubbard_chain_rule} becomes
\begin{equation}
\varepsilon^{-1} = \Omega + f\chi \Longrightarrow f = \left(\varepsilon^{-1} - \Omega\right)\chi^{-1}.
\label{eqn:deriv_of_fHxc}
\end{equation}
Finally, $U$ can be equated with the projected Hartree-plus-exchange-correlation kernel, with the residual bath screening in the background.\cite{Himmetoglu2014a} This yields
\begin{align}
 U^I = \left[\left(\frac{dv_\mathrm{KS}}{dv_\mathrm{ext}} - 1\right)\left(\frac{dn}{dv_\mathrm{ext}}\right)^{-1}\right]_{II}.
\label{eqn:variational_Dyson_equation}
\end{align}
From hereon in, we will assume that $\Omega = \delta_{IJ}$. When Hubbard projectors from different atoms overlap this may become an approximation. We will also reserve $f$ for the matrix measured via linear response, and $U$ for the parameter to be subsequently used in a DFT\,+\,$U$ calculation. This distinction will become important.

Equation \ref{eqn:variational_Dyson_equation} is nothing more than a reformulation of Eq.~\ref{eqn:U_linear_response_UminusU0}. We can identify the interacting and non-interacting response matrices
\begin{align}
   \chi_{IJ} =& \frac{dn^I}{dv^J_\mathrm{ext}}; \\
   \left(\chi_0\right)_{IJ}  =& \left[\frac{dn}{dv_\mathrm{ext}} \left(\frac{dv_\mathrm{KS}}{dv_\mathrm{ext}}\right)^{-1}\right]_{IJ}.
   \label{eqn:variational_chi0}
\end{align}
In this framework, we can see that the removal of the non-interacting response can be rigorously justified as a consequence of the Dyson equation, with $U$ being a measure of net interaction.

These definitions \edit{are nothing but a special case of} standard linear response theory for DFT (Appendix \ref{sec:chi0_in_detail}). It is crucial that the non-interacting response is calculated as the product of $\chi$ and $\varepsilon$, rather than $dn^I/dv^J_\mathrm{KS}$ directly. $dn^I / dv^J_\mathrm{KS}$ is both conceptually and numerically arbitrary with respect to the choice of external potential, and so its direct use must be circumvented.

\begin{figure}[t!]
   \includegraphics[width=3in]{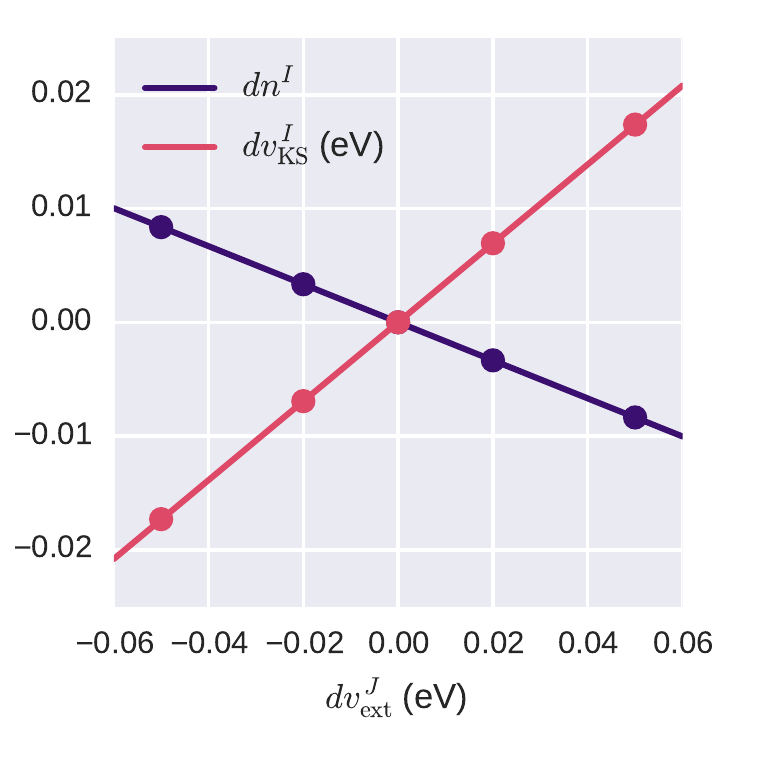}
   \caption{A typical linear response plot. Each pair of points represents an individual DFT calculation with a perturbing potential $\delta \hat v_\mathrm{ext} = dv^J_\mathrm{ext} \hat P^J$, and the resulting response of the projected density $dn^I$ and Kohn-Sham potential $dv^I_\mathrm{KS}$ . The slopes of these lines correspond to entries of $\chi$ and $\epsilon^{-1}$. These data have been taken from calculations on [Cr(H\textsubscript{2}O)\textsubscript{6}]\textsuperscript{3+}, which is covered in detail in Section \ref{sec:application}.}
   \label{fig:example_LR_plot}
\end{figure}

Figure~\ref{fig:example_LR_plot} demonstrates the calculation of elements of $\chi$ and $\varepsilon^{-1}$ from a typical set of linear response calculations.

Both minimum-tracking and conventional SCF linear response rely on the same external perturbation, and both make use of the Dyson equation. They only differ in their definition of the non-interacting response and the set of densities used in its calculation. In the minimum-tracking procedure, $\chi_0$ is constructed from ground-state densities of the perturbed system, and thus the resulting $U$ is strictly a ground-state property. This is \edit{obviously} not the case for the SCF approach; there, $\chi_0$ is calculated in reference to an unconverged density and thus the resulting $U$ is \edit{not a local property of the ground-state density landscape} (but still is a well-defined property of the ground-state Kohn-Sham eigen-system). \edit{This distinction is intriguing and worthy of further investigation, and possibly numerically inconsequential in practice.}

Already, the minimum-tracking construction reveals an interesting property of the projected $\chi_0$ (and hence $U$): it is not necessarily symmetric. This is because $\chi_0$ as defined in Eq.~\ref{eqn:variational_chi0} incorporates the total derivative of the potential, which is itself a partial derivative. While the bare $\chi_0$ is certainly symmetric, the response matrices that we deal with here are always screened by the background, and the screening depends on the subspace being perturbed. \edit{(In general, $\chi_0$ should not be symmetrized before inversion, even if the resulting $U$ matrix will be.)} This observation will also hold for SCF linear response, since it also correctly goes beyond the symmetric result of first-order perturbation theory.

%%%%%%%%%%%%%%%%%%%%%%%%%%%%%%%%%%%%%%%%%%%%%%%%%%%%%%%%%%%%%%%%%%%%%%%%%%%%%%%
\subsection{\label{subsec:spin_complications} Accounting for spin}
%%%%%%%%%%%%%%%%%%%%%%%%%%%%%%%%%%%%%%%%%%%%%%%%%%%%%%%%%%%%%%%%%%%%%%%%%%%%%%%
In the Hubbard energy functional (Eq.~\ref{eqn:DFT+U_energy}) spin and sites are treated on the same footing, with the corresponding indices being totally interchangeable. This raises the question: what happens to the response and interaction parameters if we further fine-grain linear response down to the level of spin?

In the minimum-tracking formulation it is straightforward to consider spin degrees of freedom. Response matrices become rank-four tensors
\begin{equation}
\chi^{\sigma \sigma'}_{IJ} = \frac{d n^{I\sigma}}{d v^{J\sigma'}},
\end{equation}
and to measure these elements via linear response, we must perturb spin channels individually. (Practically, this is implemented as a combination of two potentials: a uniform shift applied to both spin-channels and a spin-splitting potential.)
%Given that for closed-shell systems we would always perturb each site in turn, it is natural to perturb each spin-channel separately when performing open-shell calculations. \textcolor{red}{This is a little laboured --- if we want $\chi^{\sigma\sigma'}$ these are necessarily the calculations you have to do. Consider cutting at least part of this. I guess the argument for keeping it would be if we think the community needs comforting that we're not doing anything crazy}

This extension has several consequences. Spin-specific response functions can be visualized by flattening rank-four tensors down to rank-two ones: for example, a two-site system would have response matrices of the form
\begin{equation}
\chi =
 \begin{pmatrix}
       {\chi^{\uparrow\uparrow}_{11}}
     & {\chi^{\uparrow\downarrow}_{11}}
     & {\chi^{\uparrow\uparrow}_{12}}
     & {\chi^{\uparrow\downarrow}_{12}} \\
       {\chi^{\downarrow\uparrow}_{11}}
     & {\chi^{\downarrow\downarrow}_{11}}
     & {\chi^{\downarrow\uparrow}_{12}}
     & {\chi^{\uparrow\downarrow}_{12}} \\
       {\chi^{\uparrow\uparrow}_{21}}
     & {\chi^{\uparrow\downarrow}_{21}}
     & {\chi^{\uparrow\uparrow}_{22}}
     & {\chi^{\uparrow\downarrow}_{22}} \\
       {\chi^{\downarrow\uparrow}_{21}}
     & {\chi^{\downarrow\downarrow}_{21}}
     & {\chi^{\downarrow\uparrow}_{22}}
     & {\chi^{\downarrow\downarrow}_{22}} \\
 \end{pmatrix}
=
 \begin{pmatrix}
  {(\chi^{\sigma\sigma'})_{11}}   & {(\chi^{\sigma\sigma'})_{12}} \\
  {(\chi^{\sigma\sigma'})_{21}}   & {(\chi^{\sigma\sigma'})_{22}} \\
 \end{pmatrix}.
\end{equation}
This is not simply aesthetic: it means we are treating spin and atom indices on the same footing, like the DFT\,+\,$U$ functional does.

We can construct different models based on how we perform the inversion of this matrix (such as in Eq.~\ref{eqn:variational_Dyson_equation}): either (1) point-wise inversion, which decouples both sites and spin; (2) atom-wise inversion, with each $2 \times 2$ block inverted individually, decoupling sites but not spins; or (3) invert the full matrix, leaving all sites and spins coupled. We will work through each of \edit{them} in turn.

% \begin{center}
% \textcolor{gray}{
% \textit{The following is based on David's emails on 27 Dec and 27 Jan}
% }
% \end{center}
\begin{figure*}[t]
   \includegraphics[width=\textwidth]{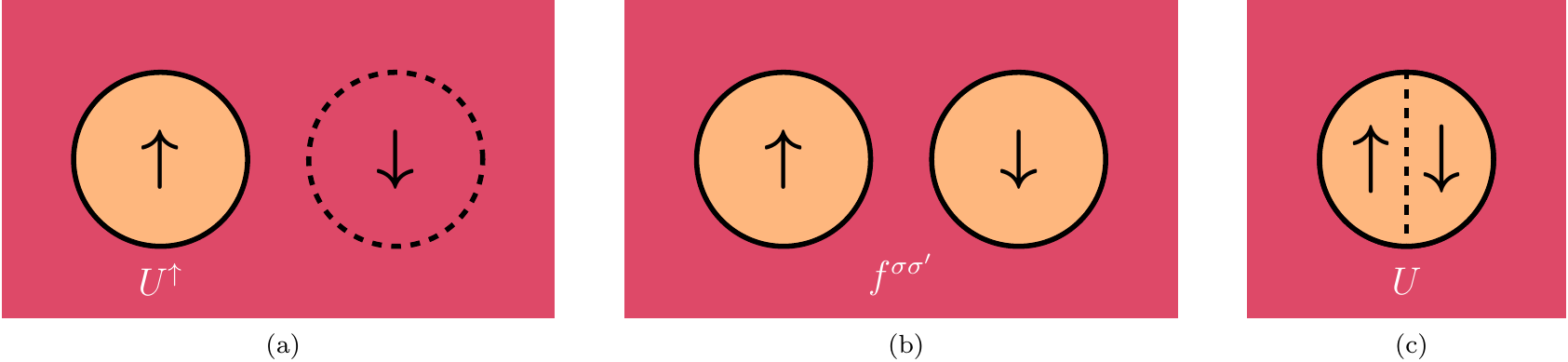}
   \caption{Schematic diagram illustrating which subspaces screen the Hubbard parameters (pink) and which do not (orange). Individual Hubbard sites are represented by solid circles. Point-wise inversion (a) effectively treats our system as a one-site Hubbard model connected to a bath, where the bath includes the opposite-spin subspace of the same site. Atom-wise inversion (b) is effectively a two-site system connected to a bath. Finally, in conventional linear response (c) both spin channels on a given atom are treated as a single Hubbard site.}
\label{fig:cartoon_systems}
\end{figure*}

\subsubsection{Point-wise inversion}
% \begin{figure}[h!]
%    \centering
%    \input{fig_cartoon_1x1.tex}
%    \caption{Schematic diagram illustrating which subspaces screen the Hubbard parameters (pink) and which do not (orange). Individual Hubbard sites are represented by solid circles. Point-wise inversion effectively treats our system as a one-site Hubbard model connected to a bath, where the bath includes the opposite-spin subspace of the same site. %Wherever a subspace is perturbed with respect to another, the change in relative energy between them introduces screening in the resulting response matrix element.
% In this case, $U^\uparrow$ is screened by the corresponding spin-down subspace and the rest of the system.}
% % \label{fig:cartoon_systems}
%    \label{fig:cartoon_1x1}
% \end{figure}
The Hubbard parameters in this case are screened by the opposite spin on the same site (Fig.~\ref{fig:cartoon_systems}a). In this case, equation~\ref{eqn:variational_Dyson_equation} separates into an independent equation for each atom:
\begin{equation}
f^{\sigma \sigma} = \frac{d v^\sigma_\mathrm{KS}}{d n^{\sigma}} - \frac{d v^\sigma_\mathrm{ext}}{d n^{\sigma}} = \frac{d v^{\sigma}_\mathrm{Hxc}}{d n^{\sigma}}.
\end{equation}
We have dropped the atomic indices for brevity. This simplification affords some numerical cancellation of errors, since inversion is no longer performed. The off-diagonal components of the matrix $f^{\sigma \sigma'}$ are not meaningful in this case. The conventional DFT\,+\,$U$ functional requires a spin-independent $U$; for this we must average the spin-up and spin-down components:
\begin{equation}
U = \frac{1}{2}\left(f^{\uparrow \uparrow} + f^{\downarrow \downarrow}\right).
\end{equation}
This will henceforth be referred to as ``averaged 1$\times$1". There is also the option to avoid this approximation and apply a different value of $U$ to each spin channel: $U^\sigma$ = $f^{\sigma \sigma}$ (``1$\times$1").

\edit{It is interesting to note that Shishkin and Sato\cite{Shishkin2016} have previously advocated removing the off-diagonal components of site-indexed response matrices. This was motivated by the fact that these components were negligible so removing them did not alter the resulting Hubbard parameters. Here, however, the off-diagonal components components correspond to coupling between spin channels on the same atom. These components are sizeable and neglecting them appreciably alters Hubbard and Hund's parameters, as we will see.}

\subsubsection{Atom-wise inversion}
% \begin{figure}[h!]
%    \input{fig_cartoon_2x2.tex}
%    \caption{Atom-wise inversion is effectively a two-site system connected to a bath. While screening from both the bath and other sites is present in the response matrices, there is no inter-site screening in $f^{\sigma\sigma'}$ as it is removed by the inversion of the spin-indexed response.}
%    \label{fig:cartoon_2x2}
% \end{figure}
In atom-wise inversion, screening from both the bath and other sites is present in the response matrices, but the resulting $f = \chi_0^{-1} - \chi^{-1}$ is \emph{bare} with respect to inter-spin interactions on the same atom as it is removed by the inversion of the spin-indexed response (Fig.~\ref{fig:cartoon_systems}b). Employing this approach amounts to assuming inter-spin interactions will be corrected separately \emph{i.e.}\ with a +\,$J$ functional. (This is because in the absence of such a correction, a spin-screened $U$ would be necessary.)

Equation~\ref{eqn:variational_Dyson_equation} reduces to
\begin{equation}
f^{\sigma\sigma'} = \left[\left(\frac{d {v_\mathrm{KS}}}{d {v_\mathrm{ext}}} - 1\right) \left(\frac{d n}{d {v_\mathrm{ext}}}\right)^{-1}\right]^{\sigma\sigma'}
\end{equation}
where each term is a two-by-two matrix indexed by spin channel, and if there are $N$ atoms there are $N$ such equations. For practical use in DFT\,+\,$U$\,+\,$J$, $f$ can be related to the scalar Hubbard parameter $U$ \edit{that, in the minimum-tracking linear-response formalism, is defined by}
\begin{align}
   U =& \frac{1}{2}\frac{dv_\mathrm{Hxc}^\uparrow + dv_\mathrm{Hxc}^{\downarrow}}{d(n^{\uparrow} + n^{\downarrow})} \nline
   \approx& \frac{1}{2}\frac{f^{\uparrow\uparrow}\delta n^\uparrow +f^{\uparrow\downarrow}\delta n^\downarrow +f^{\downarrow\uparrow} \delta n^\uparrow +f^{\downarrow\downarrow}\delta n^\downarrow}{\delta (n^{\uparrow} + n^{\downarrow})}.
   \label{eqn:U_as_an_exact_frac}
\end{align}
There are two alternative approximations we can make here. The first, more na\"ive approach, is to further approximate this as
\begin{equation}
   U = \frac{1}{4}\left(f^{\uparrow \uparrow} + f^{\uparrow \downarrow} + f^{\downarrow \uparrow} + f^{\downarrow \downarrow}\right)
   % U = \frac{1}{4}\sum_{\sigma \sigma'} f^{\sigma \sigma'}
   \label{eqn:U_2x2a}
\end{equation}
which we will refer to as ``simple 2$\times$2". A more sophisticated approach (scaled ``$2\times2$") is
\begin{align}
U =& \frac{1}{2}
   \frac{
      \lambda_U(
         f^{\uparrow\uparrow}
         + f^{\downarrow\uparrow}
         ) 
      + f^{\uparrow\downarrow}
      + f^{\downarrow\downarrow}
      }{
         \lambda_U + 1
      }; \label{eqn:U_2x2b}\\
\label{eqn:lambda_U_2x2b}
   \lambda_U 
   =& \frac{
      \chi^{\uparrow\uparrow} + \chi^{\uparrow \downarrow}
      }{
      \chi^{\downarrow \uparrow} + \chi^{\downarrow \downarrow}
      }.
\end{align}
%
%
%In the case of a closed shell system, this simplifies due to the fact that $\delta n^\uparrow = \delta n^\downarrow$, $f^{\uparrow\uparrow} = f^{\downarrow\downarrow}$,  and $f^{\uparrow\downarrow} = f^{\downarrow\uparrow}$:
%%
%\begin{equation}
%U = \frac{1}{2}(f^{\uparrow\uparrow} + f^{\uparrow\downarrow})
%\end{equation}
%%
%that is, $U$ is the average of the like- and unlike-spin interactions. Therefore, a sensible (but approximate) averaging scheme for open-shell systems is
%%
%\begin{equation}
%U = \frac{1}{4}\left(f^{\uparrow \uparrow} + f^{\uparrow \downarrow} + f^{\downarrow \uparrow} + f^{\downarrow \downarrow}\right)
%   \end{equation}
%
%    \begin{equation}
% J = -\frac{1}{4}\left(f^{\uparrow \uparrow} - f^{\uparrow \downarrow} - f^{\downarrow \uparrow} + f^{\downarrow \downarrow}\right)
%    \end{equation}
% %
The derivations of \cref{eqn:U_2x2a,eqn:U_2x2b} involve varying levels of approximation, which are discussed in detail in Appendix~\ref{sec:the_lambda_approximation}.

With atom-wise inversion, Hund's parameters $J$ can be directly calculated in an analogous manner to $U$: in place of Eq.~\ref{eqn:U_as_an_exact_frac} we instead \edit{define, within the spin-polarized minimum-tracking linear response formalism,}
\begin{equation}
   J = -\frac{1}{2}\frac{dv_\mathrm{Hxc}^\uparrow - dv_\mathrm{Hxc}^{\downarrow}}{d(n^{\uparrow} - n^{\downarrow})}
   \label{eqn:J_exact}
\end{equation}
%
% where $\mu \equiv n^\uparrow - n^\downarrow$.
%
\edit{For $J$}, simple $2\times2$ yields
\begin{equation}
   J = -\frac{1}{4}\left(f^{\uparrow \uparrow} - f^{\uparrow \downarrow} - f^{\downarrow \uparrow} + f^{\downarrow \downarrow}\right),
   \label{eqn:J_2x2a}
\end{equation}
while scaled $2\times2$ gives
\begin{align}
J =& -\frac{1}{2}\frac{\lambda_J(f^{\uparrow\uparrow} - f^{\downarrow\uparrow}) + f^{\uparrow\downarrow} - f^{\downarrow\downarrow}}{\lambda_J - 1}
\label{eqn:J_2x2b}; \\
\lambda_J =& \frac{\chi^{\uparrow\uparrow} - \chi^{\uparrow \downarrow}}{\chi^{\downarrow \uparrow} - \chi^{\downarrow \downarrow}}.
\end{align}
%The averaging over spin channels removes any asymmetry that the Hubbard parameters would otherwise inherit from $\chi_0$ (which was discussed earlier).
%
% Note that the inversion of the spin-indexed matrices rigorously decouples the response of $E$ with respect to $n$ and $\mu$. This does not happen in conventional approaches for calculating $U$ and $J$ \cite{Himmetoglu2011a}, where the calculations of $d^2E/dn^2$ and $d^2E/d\mu^2$ are performed independently and do account for changes in $\mu$ and $n$ respectively. This conventional approach risks conflating the two curvatures.

\subsubsection{Full inversion}
Finally, in the case of full matrix inversion, the result is bare with respect to both inter-spin and inter-site interactions by the same logic. This implies that inter-atom interactions require, and are subject to, correction via a +\,$V$ term. This $V$ term would be doubly spin-dependent, and it may need to be symmetrized with respect to the site indices to retain a Hermitian Kohn-Sham Hamiltonian for each spin. We will not explore this approach further in this work.

We emphasize that including each of these successive terms ($J$ and $V$) should not be viewed as systematic improvements. In the limit that corrective parameters are introduced within and between every single subspace (such that the corresponding \edit{screened interactions} are removed) the entire system becomes effectively non-interacting. Corrective terms are only appropriate where the corresponding interactions dwarf all others.

%In theory, by comparing the Hubbard parameter values as given by these three different approaches, one could systematically determine when $+J$ and $+V$ corrections are important for a given system. \textcolor{red}{We need to be very clear on this front as we want to subsequently use method 1a for systems where 2 gives very different values. Perhaps delay this discussion until when discussing a good example in the hexahydrated metal dataset?}

%%%%%%%%%%%%%%%%%%%%%%%%%%%%%%%%%%%%%%%%%%%%%%%%%%%%%%%%%%%%%%%%%%%%%%%%%%%%%%%
\subsection{Comparisons with the conventional scalar approach}
%%%%%%%%%%%%%%%%%%%%%%%%%%%%%%%%%%%%%%%%%%%%%%%%%%%%%%%%%%%%%%%%%%%%%%%%%%%%%%%
\label{subsec:comparisons_with_conventional_approach}

% \begin{figure}[h!]
%    \input{fig_cartoon_conv.tex}
%    \caption{In conventional linear response, both spin channels on a given atom are treated as a single Hubbard site.}
%    \label{fig:cartoon_conv}
% \end{figure}

Conventional linear response calculations do not treat spin channels separately (Fig.~\ref{fig:cartoon_systems}c); for a single-site system $\chi$, $\varepsilon^{-1}$ and $f$ would all be scalars. It is straightforward to relate the spin-indexed response matrices of the previous section to these scalars:
\begin{align}
d n      = d n^\uparrow + d n^{\downarrow} 
              \approx & \left[\sum_{\sigma\sigma'} \chi^{\sigma\sigma'}\right]d v_\mathrm{ext} \nline
\Longrightarrow \chi \approx & \sum_{\sigma\sigma'} \chi^{\sigma\sigma'}.
\label{eqn:chi_relation}
\end{align}
Likewise
\begin{align}
dv_\mathrm{KS} = \frac{1}{2}\left[d v^\uparrow_\mathrm{KS} + d v^{\downarrow}_\mathrm{KS}\right] \approx & \frac{1}{2}\left[\sum_{\sigma\sigma'} \left(\varepsilon^{-1}\right)^{\sigma\sigma'}\right]d v_\mathrm{ext} \nline
\Longrightarrow \varepsilon^{-1} \approx & \frac{1}{2}\left[\sum_{\sigma\sigma'} \left(\varepsilon^{-1}\right)^{\sigma\sigma'}\right].
\label{eqn:varepsilon_relation}
\end{align}
These two relations allow us to examine the role of spin-screening in scalar linear response. The Hubbard parameter obtained via spin-indexed, atom-wise inversion (scaled $2\times2$; Eq.~\ref{eqn:U_2x2b}) can be rewritten as
\begin{align}
U
   =& \frac{1}{2}
      \frac{
         \sum_{\sigma \sigma'} (f\chi)^{\sigma \sigma'}
         }{
            \sum_{\sigma \sigma'} \chi^{\sigma \sigma'}
         } \nline
   =& \frac{1}{2}
      \frac{
         \sum_{\sigma \sigma'} (\varepsilon^{-1} - 1)^{\sigma \sigma'}
         }{
            \sum_{\sigma \sigma'} \chi^{\sigma \sigma'}
         } \nline
   =& \frac{\varepsilon^{-1} - 1}{\chi}.
\end{align}
This is nothing less than the scalar expression $U = \chi_0^{-1} - \chi^{-1}$, which is used in scalar linear response. We may conclude that the conventional scalar approach and scaled $2\times2$ are entirely equivalent.

Therefore, Hubbard parameters obtained by \edit{spin-aggregated} approaches are not screened by the opposite spin channel on the same site. \edit{Since they combine both like and unlike spin interactions (\emph{c.f.}\ Eq.~\ref{eqn:U_2x2b}), they} do \emph{not} correspond to the \edit{like-spin-only interaction} $U_\text{eff} = U - J$ (as implied elsewhere).\cite{Kulik2008a} We could have anticipated this result: during a scalar linear response calculation there is no shift in the external potential difference between the two spin channels, so (to first order) there is no external driver for changes in subspace spin polarization. % Consequently, there is no charge transfer between them (to first order) and hence there is no screening by the opposite spin. 

We noted earlier that atom-wise inversion formally necessitates a Hund's correction, but such a correction is not usually included when the conventional linear response approach is employed. Given that these methods are equivalent, we argue that it is more consistent to include a Hund's exchange correction term \edit{(\emph{e.g.}\ calculated using Eq.~\ref{eqn:J_exact})} if using a Hubbard correction calculated in the conventional manner.

\edit{The precise functional form of the +\,$J$ correction needed is, however, the subject of ongoing research. Recently, for example, Millis and co-workers demonstrated that spin-polarized DFT already possesses some degree of intrinsic exchange splitting, and they have argued convincingly that the contemporary form of the +\,$J$ correction can overestimate exchange splitting.\cite{Chen2016} This finding is corroborated by our own results discussed later in this paper (\emph{e.g.}\ Table~\ref{tab:ddExcitations_spin_flips}).}

% \edit{This requires caution: Millis and co-workers have demonstrated that spin-polarized DFT already possesses some degree of intrinsic exchange splitting, so a $+J$ correction can overestimate exchange splitting.\cite{Chen2016} (This proposal is corroborated by results later in this paper.) The precise functional form of the $+J$ correction is the subject of ongoing research.}
% \textcolor{red}{I'm not sure if/where to include the result of my email 24 Aug 2017}

%%%%%%%%%%%%%%%%%%%%%%%%%%%%%%%%%%%%%%%%%%%%%%%%%%%%%%%%%%%%%%%%%%%%%%%%%%%%%%%
\section{\label{sec:application} Application to a complete series of hexahydrated transition metals and manganese oxide}
%%%%%%%%%%%%%%%%%%%%%%%%%%%%%%%%%%%%%%%%%%%%%%%%%%%%%%%%%%%%%%%%%%%%%%%%%%%%%%%

\begin{figure}[t!]
   \includegraphics[width=\columnwidth]{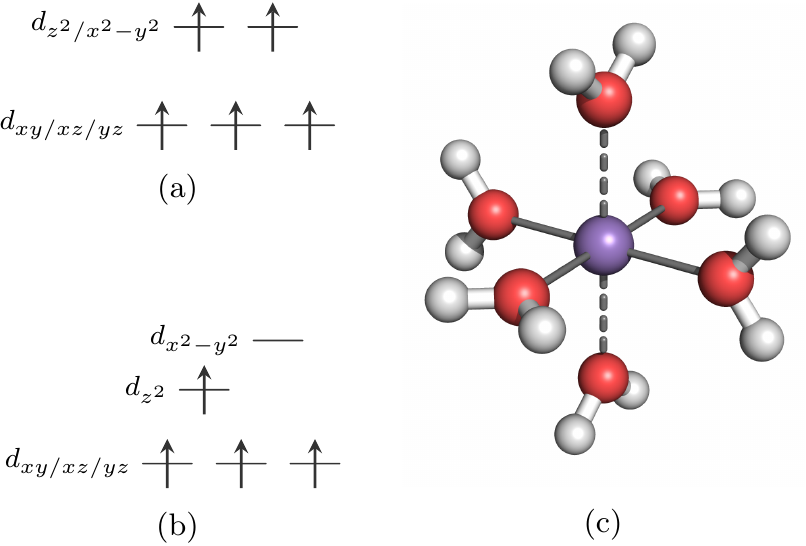}
   \caption{ The ground state of the $3d$ electrons in (a) [Mn(H$_2$O)$_6$]$^{2+}$ and (b) [Mn(H$_2$O)$_6$]$^{3+}$. In both systems, the $d_{xy}$, $d_{xz}$, and $d_{yz}$ orbitals have lower energy as they have lobes directed between the ligands (and hence less overlap with the ligand orbitals). For the doubly-charged system, the system is symmetric and no Jahn-Teller splitting takes place. In the triply-charged system, the molecule distorts into a $D_{2h}$ symmetry as shown in (c), with the axial bonds (dashed) fractionally longer than the equatorial bonds (solid).}
   \label{fig:hexMn}
\end{figure}
In the second half of this work, we explore the ramifications of our theoretical developments on \edit{two test systems: hexahydrated transition metals, and manganese oxide.}

In these systems, all of the metal atoms have partially filled $3d$ sub-shells. Electrons within these sub-shells are in such close proximity to one another that the interplay of their spin, charge, and orbital moment are too pronounced to be well described by local or semi-local xc-functionals.\cite{Harvey2006a,Cramer2009a,Yu2016a,Johnson2017a} DFT\,+\,$U$ may provide a more accurate description of these systems.\cite{Kulik2010a,Kulik2011a,Himmetoglu2014a,Zhao2016a}

\edit{Manganese oxide (MnO) has a rock salt structure. At low temperatures it is antiferromagnetic,\cite{Shull1951a} and has a band gap of approximately 4\,eV that is substantially underestimated by semi-local functionals.\cite{Anisimov1997a} Conventional linear-response calculations on MnO yield an excessively large Hubbard parameter ($U > 7$\,eV).\cite{Lim2016}}

\edit{Meanwhile,} hexahydrated transition metals comprise of a central first-row transition metal ion surrounded by six water ligands in a tetragonal arrangement (Fig.~\ref{fig:hexMn}c). Such systems \edit{bear some resemblance} to a fundamental unit of transition metal oxides, as well as organometallic systems such as the oxygen evolving complex of photosystem II.\cite{Umena2011a, Suga2015a}

Depending on the electronic structure of the metal, these systems may exhibit Jahn-Teller distortion, resulting in an elongated tetragonal structure with two axial waters being slightly more distant than their four equatorial counterparts (Fig.~\ref{fig:hexMn}).

%%%%%%%%%%%%%%%%%%%%%%%%%%%%%%%%%%%%%%%%%%%%%%%%%%%%%%%%%%%%%%%%%%%%%%%%%%%%%%%
\subsection{\label{subsec:computational_details} Computational details}
%%%%%%%%%%%%%%%%%%%%%%%%%%%%%%%%%%%%%%%%%%%%%%%%%%%%%%%%%%%%%%%%%%%%%%%%%%%%%%%
All calculations were performed using ONETEP\cite{Skylaris2005a, Hine2011a, ORegan2012a, ORegan2010a, Ruiz-Serrano2012a, Fox2011a, ORegan2011a} (Order-$N$ Electronic Total Energy Package, version 4.3) using the Perdew-Burke-Ernzerhof (PBE) xc-functional.\cite{Perdew1996a} 

\edit{For MnO, a square super-cell containing 512 atoms was simulated under periodic boundary conditions without explicit $k$-point sampling. This is a non-diagonal super-cell\cite{LloydWilliams2015} of the four-atom primitive cell, and gives an equivalent $k$-point sampling scheme that includes both $Z$ and $\Gamma$. (This is crucial because the band gap of MnO is known to be $Z$ to $\Gamma$.) The lattice parameter was set to the experimental value of 4.445\,\AA.\cite{Wyckoff1963a} The calculations were spin-polarized, with an energy cut-off of 1030\,eV. ONETEP uses a basis of non-orthogonal generalized Wannier functions (NGWFs) that are variationally optimized \textit{in situ}. Each Mn atom had ten NGWFs; O atoms, four. All NGWFs had a cutoff radius of 11.0\,$a_0$.}

For the hexahydrated metals, all calculations were spin-polarized, with an energy cut-off of 897\,eV. Depending on the species, there were 9, 10, or 13 NGWFs on the transition metal atom, four on each oxygen, and one on each hydrogen. All NGWFs had 14\,$a_0$ cutoff radii. An Elstner dispersion correction\cite{Elstner2001a, Hill2009a} was applied, and electrostatics were treated using a padded cell and a Coulomb cut-off.\cite{Hine2011b}

For all the calculations, the Hubbard projectors were constructed from solving the neutral atomic problem subject to the pseudopotential of the species in question.\cite{Ruiz-Serrano2012a} Most pseudopotentials were taken from the Rappe group pseudopotential library\cite{RappeGroupPsps} although those for Co and Fe were generated in-house using OPIUM.\cite{opium,Kerker1980a,Kleinman1982a,Hamann1989a,Rappe1990a,Gonze1991a,Ramer1999a} These were scalar relativistic pseudopotentials\cite{Grinberg2000a} with non-linear core corrections.\cite{Louie1982a}. All DFT\,+\,$U$\,+\,$J$ calculations used a $+J$ correction to the energy, potential, and ionic forces. We used the energetic correction shown in Eq.~\ref{eqn:DFT+J_energy} (following the example of Ref.~\onlinecite{Himmetoglu2011a} we have omitted the ``$n_\mathrm{min}$" term that appears in that paper).

Example input and output files can be found at \texttt{www.repository.cam.ac.uk/}.
%%%%%%%%%%%%%%%%%%%%%%%%%%%%%%%%%%%%%%%%%%%%%%%%%%%%%%%%%%%%%%%%%%%%%%%%%%%%%%%
\subsection{Calculating Hubbard parameters}
\label{subsec:hexX_Hubbard_parameters}
%%%%%%%%%%%%%%%%%%%%%%%%%%%%%%%%%%%%%%%%%%%%%%%%%%%%%%%%%%%%%%%%%%%%%%%%%%%%%%%

\begin{table*}[t!]
\centering
\caption{Values of $U$ and $J$ (eV) for hexahydrated transition metals \edit{and a spin-up manganese atom of MnO}, calculated using the various linear response schemes introduced in subsection~\ref{subsec:spin_complications}. The linear response calculations for the fully-filled $3d$ subspace in Fe\textsuperscript{3+} were poorly behaved (two different pseudopotentials were tested) and have consequently been excluded.}
\label{tab:Udet_metal}
\footnotesize
\begin{tabular}{m{1.5cm} |
d @{$\,\pm\,$} 
m{0.8cm} |
d @{$\,\pm\,$}
m{0.8cm} |
d @{$\,\pm\,$}
m{0.8cm}
d @{$\,\pm\,$}
m{0.8cm} |
d @{$\,\pm\,$}
m{0.8cm}
d @{$\,\pm\,$}
m{0.8cm} |
d @{$\,\pm\,$}
m{0.8cm}
d @{$\,\pm\,$}
m{0.8cm}
d @{$\,\pm\,$}
m{0.8cm}
d @{$\,\pm\,$}
m{0.8cm}}
\hline
\multicolumn{1}{c |}{\multirow{2}{*}{metal}}
&\multicolumn{2}{c |}{scalar}
&\multicolumn{2}{c |}{averaged $1\times1$}
&\multicolumn{4}{c |}{$1\times1$}
&\multicolumn{4}{c |}{simple $2\times2$}
&\multicolumn{4}{c }{scaled $2\times2$} \\
&\multicolumn{2}{c|}{$U$}
&\multicolumn{2}{c|}{$U$}
&\multicolumn{2}{c}{$U^\uparrow$}
&\multicolumn{2}{c|}{$U^\downarrow$}
&\multicolumn{2}{c}{$U$}
&\multicolumn{2}{c|}{$J$}
&\multicolumn{2}{c}{$U$}
&\multicolumn{2}{c}{$J$} \\
\hline
\centering Ti$^{3+}$ & 3.88 & 0.00 & 1.66 & 0.00 & 1.85 & 0.01 & 1.47 & 0.00 & 3.90 & 0.01 & 0.34 & 0.00 & 3.89 & 0.01 & 0.34 & 0.00 \\
\centering V$^{2+}$ & 4.00 & 0.00 & 2.78 & 0.00 & 3.29 & 0.00 & 2.28 & 0.00 & 4.07 & 0.01 & 0.34 & 0.00 & 4.00 & 0.01 & 0.35 & 0.00 \\
\centering Cr$^{3+}$ & 3.90 & 0.00 & 1.78 & 0.00 & 1.86 & 0.00 & 1.70 & 0.00 & 4.04 & 0.01 & 0.40 & 0.00 & 3.90 & 0.01 & 0.42 & 0.00 \\
\centering Cr$^{2+}$ & 3.20 & 0.00 & 2.39 & 0.00 & 2.75 & 0.00 & 2.04 & 0.00 & 3.34 & 0.01 & 0.33 & 0.00 & 3.20 & 0.01 & 0.35 & 0.00 \\
\centering Mn$^{3+}$ & 5.40 & 0.00 & 2.00 & 0.00 & 1.51 & 0.00 & 2.50 & 0.00 & 5.86 & 0.01 & 0.50 & 0.00 & 5.40 & 0.01 & 0.53 & 0.00 \\
\centering Mn$^{2+}$ & 4.36 & 0.00 & 4.05 & 0.08 & 4.28 & 0.15 & 3.82 & 0.00 & 4.90 & 0.06 & 0.37 & 0.06 & 4.35 & 0.01 & 0.52 & 0.01 \\
\centering Fe$^{3+}$ & 5.88 & 0.01 & \multicolumn{2}{c|}{---} & \multicolumn{2}{c}{---} & 5.45 & 0.02 & \multicolumn{2}{c}{---} & \multicolumn{2}{c|}{---} & 5.92 & 0.02 & 0.81 & 0.02 \\
\centering Fe$^{2+}$ & 4.58 & 0.00 & 5.07 & 0.09 & 6.28 & 0.18 & 3.86 & 0.00 & 6.06 & 0.09 & 0.43 & 0.06 & 4.58 & 0.01 & 0.63 & 0.01 \\
\centering Co$^{3+}$ & 6.25 & 0.00 & 1.19 & 0.00 & 1.19 & 0.00 & 1.19 & 0.00 & 6.25 & 0.00 & 0.75 & 0.00 & 6.25 & 0.00 & 0.75 & 0.00 \\
\centering Co$^{2+}$ & 4.95 & 0.02 & 6.19 & 0.02 & 8.17 & 0.03 & 4.22 & 0.02 & 7.15 & 0.02 & 0.48 & 0.01 & 4.96 & 0.02 & 0.65 & 0.01 \\
\centering Ni$^{2+}$ & 5.26 & 0.00 & 9.84 & 0.02 & 15.41 & 0.05 & 4.27 & 0.00 & 12.35 & 0.03 & 0.75 & 0.02 & 5.26 & 0.01 & 0.78 & 0.01 \\
\centering Cu$^{2+}$ & 4.62 & 0.00 & -2.54 & 0.03 & -9.11 & 0.05 & 4.04 & 0.00 & -4.99 & 0.02 & 0.85 & 0.02 & 4.63 & 0.01 & 0.90 & 0.01 \\

\hline
\centering MnO & 5.44 & 0.04 & 4.63 & 0.08 & 5.54 & 0.15 & 3.72 & 0.02 & 8.38 & 0.15 & 0.51 & 0.05 & 5.37 & 0.04 & 0.49 & 0.02 \\

\hline
\end{tabular}

\vspace{2ex}
\caption{Values of $U$ and $J$ (eV) \edit{calculated using the various linear response schemes, for an equatorial oxygen atom within hexahydrated transition metal systems, and for a MnO oxygen atom.}}
\label{tab:Udet_oxygen}
\footnotesize
\begin{tabular}{m{1.5cm} |
d @{$\,\pm\,$} 
m{0.8cm} |
d @{$\,\pm\,$}
m{0.8cm} |
d @{$\,\pm\,$}
m{0.8cm}
d @{$\,\pm\,$}
m{0.8cm} |
d @{$\,\pm\,$}
m{0.8cm}
d @{$\,\pm\,$}
m{0.8cm} |
d @{$\,\pm\,$}
m{0.8cm}
d @{$\,\pm\,$}
m{0.8cm}
d @{$\,\pm\,$}
m{0.8cm}
d @{$\,\pm\,$}
m{0.8cm}}
\hline
\multicolumn{1}{c |}{\multirow{2}{*}{metal}}
&\multicolumn{2}{c |}{scalar}
&\multicolumn{2}{c |}{averaged $1\times1$}
&\multicolumn{4}{c |}{$1\times1$}
&\multicolumn{4}{c |}{simple $2\times2$}
&\multicolumn{4}{c }{scaled $2\times2$} \\
&\multicolumn{2}{c|}{$U$}
&\multicolumn{2}{c|}{$U$}
&\multicolumn{2}{c}{$U^\uparrow$}
&\multicolumn{2}{c|}{$U^\downarrow$}
&\multicolumn{2}{c}{$U$}
&\multicolumn{2}{c|}{$J$}
&\multicolumn{2}{c}{$U$}
&\multicolumn{2}{c}{$J$} \\
\hline
\centering Ti$^{3+}$ & 8.16 & 0.03 & 5.05 & 0.01 & 5.20 & 0.01 & 4.89 & 0.00 & 8.14 & 0.02 & 1.05 & 0.00 & 8.13 & 0.02 & 1.05 & 0.00 \\
\centering V$^{2+}$ & 8.28 & 0.00 & 5.69 & 0.00 & 5.70 & 0.00 & 5.69 & 0.00 & 8.28 & 0.01 & 1.29 & 0.00 & 8.28 & 0.01 & 1.29 & 0.00 \\
\centering Cr$^{3+}$ & 8.29 & 0.00 & 5.54 & 0.00 & 5.44 & 0.00 & 5.65 & 0.00 & 8.29 & 0.02 & 1.08 & 0.01 & 8.29 & 0.02 & 1.08 & 0.01 \\
\centering Cr$^{2+}$ & 8.44 & 0.01 & 6.28 & 0.01 & 6.55 & 0.01 & 6.01 & 0.02 & 8.45 & 0.02 & 1.27 & 0.01 & 8.45 & 0.02 & 1.27 & 0.01 \\
\centering Mn$^{3+}$ & 8.57 & 0.00 & 4.94 & 0.00 & 5.53 & 0.00 & 4.35 & 0.00 & 8.58 & 0.03 & 0.97 & 0.01 & 8.57 & 0.03 & 0.97 & 0.01 \\
\centering Mn$^{2+}$ & 8.30 & 0.00 & 6.05 & 0.01 & 5.70 & 0.01 & 6.39 & 0.00 & 8.29 & 0.01 & 1.31 & 0.00 & 8.31 & 0.01 & 1.30 & 0.01 \\
\centering Fe$^{3+}$ & 8.37 & 0.03 & 5.55 & 0.05 & 4.48 & 0.07 & 6.62 & 0.07 & 8.59 & 0.14 & 1.24 & 0.06 & 8.40 & 0.12 & 1.06 & 0.06 \\
\centering Fe$^{2+}$ & 8.83 & 0.01 & 5.77 & 0.00 & 5.43 & 0.01 & 6.10 & 0.00 & 8.83 & 0.01 & 1.40 & 0.00 & 8.83 & 0.01 & 1.39 & 0.01 \\
\centering Co$^{3+}$ & 8.26 & 0.00 & 4.37 & 0.09 & 4.27 & 0.11 & 4.48 & 0.15 & 8.39 & 0.10 & 1.12 & 0.05 & 8.39 & 0.10 & 1.12 & 0.05 \\
\centering Co$^{2+}$ & 8.25 & 0.06 & 5.24 & 0.10 & 4.89 & 0.11 & 5.60 & 0.15 & 8.24 & 0.09 & 1.38 & 0.06 & 8.25 & 0.09 & 1.37 & 0.06 \\
\centering Ni$^{2+}$ & 8.09 & 0.01 & 4.89 & 0.00 & 4.65 & 0.00 & 5.14 & 0.00 & 8.09 & 0.01 & 1.37 & 0.00 & 8.09 & 0.01 & 1.37 & 0.00 \\
\centering Cu$^{2+}$ & 8.38 & 0.00 & 5.08 & 0.00 & 4.68 & 0.00 & 5.48 & 0.00 & 8.36 & 0.01 & 1.38 & 0.00 & 8.38 & 0.01 & 1.38 & 0.00 \\

\hline
\centering MnO & 10.88 & 0.01 & 5.32 & 0.04 & 5.32 & 0.05 & 5.32 & 0.05 & 10.92 & 0.12 & 1.03 & 0.03 & 10.92 & 0.12 & 1.03 & 0.03 \\

\hline
\end{tabular}
\end{table*}

\begin{figure*}[t!]
   \includegraphics[width=7in]{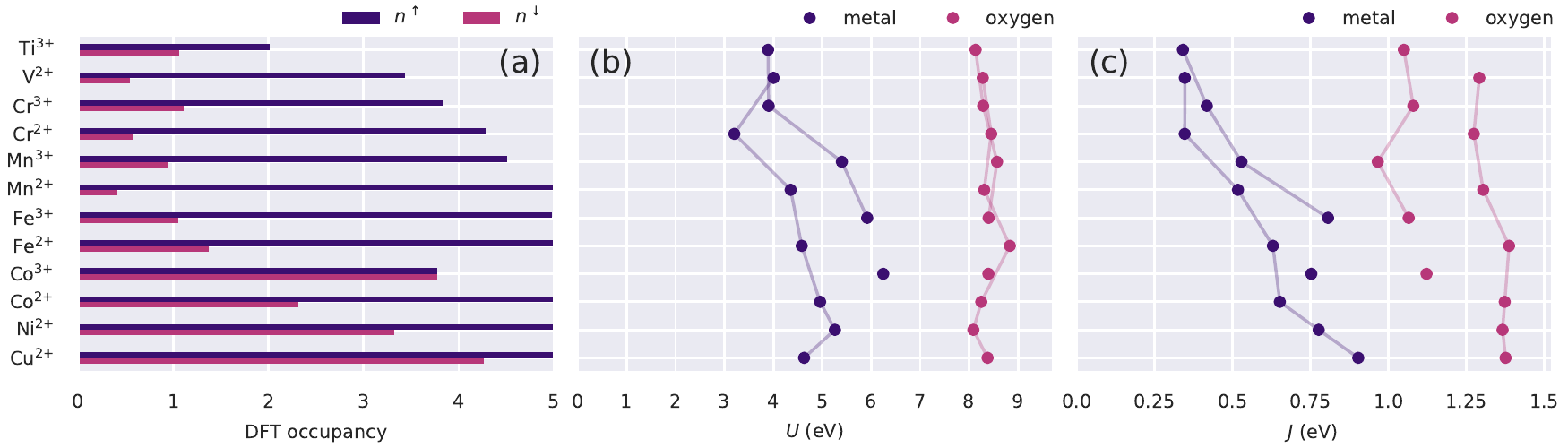}
   \caption{\edit{(a) The metal ion 3$d$ subspace occupancies as given by DFT. The residual spin-down densities for the lighter metals are not formally what one would expect; a Hubbard correction should remedy this. (b) Hubbard parameters and (c) Hund's parameters as calculated via scaled $2\times2$ (equivalent to the scalar approach). Faint lines link the $+2$ systems/$+3$ systems to show the general trends. (Co\textsuperscript{3+}, being the only low-spin system, is not linked.)}}
   \label{fig:combined_U_J_and_occ}
\end{figure*}
% \begin{figure*}[t!]
%    \includegraphics[width=7in]{fig_J_A_2_values_plot.pdf}
%    \caption{Hund's parameters as calculated via scaled $2\times2$.}
%    \label{fig:UandJ_A_2_values}
% \end{figure*}
% 
% % \begin{figure*}
% % \includegraphics[width=7in]{fig_J_A_2_values_plot.pdf}
% % \caption{$J$ as calculated via scheme 2}
% % \label{fig:J_A_2_values}
% % \end{figure*}
% 
% \begin{figure*}
% \includegraphics[width=7in]{fig_DFT_3d_occupancies_bar_plot.pdf}
% \label{fig:occupancy_bar_plot}
% \end{figure*}

\edit{Hubbard $U$ and Hund's $J$ parameters} were calculated for a set of hexahydrated transition metals. Prior to the linear response calculations, the geometries of every system were optimized using the PBE xc-functional without a Hubbard correction and with the water molecules constrained to their respective planes. Various linear response approaches were performed: averaged and non-averaged $1\times1$, simple and scaled $2\times2$, as well as the standard scalar approach. While the scalar values reported here will be roughly analogous to conventional linear response reported elsewhere, they were calculated using minimum-tracking linear response, not SCF, which differ in their definitions of $\chi_0$.

Hubbard and Hund's parameters were obtained for two Hubbard subspaces: the $3d$ subspace on the transition metal ion, and the $2p$ subspace on one of the equatorial oxygen atoms, taken as a representative of the six oxygen atoms in the system. The Hubbard parameters that were obtained are listed in Tables~\ref{tab:Udet_metal} and \ref{tab:Udet_oxygen} respectively, and plotted in Fig.~\ref{fig:combined_U_J_and_occ}. The uncertainties in the Hubbard parameters have also been calculated from the error in the least-square fits of $d v^\sigma_\text{Hxc}/d n^{\sigma'}$, $d v^\sigma_\mathrm{KS}/d v^{\sigma'}_\text{ext}$ and $d n^\sigma/d v^{\sigma'}_\text{ext}$ using unbiased Gaussian error propagation. These error estimates prove to be very instructive.

\subsubsection{General trends}
Both tables exhibit some general trends: the Hubbard parameters of the metal ions grow slowly as the number of $3d$ electrons increases (Fig.~\ref{fig:combined_U_J_and_occ}a); oxygen parameters remain relatively stable; the Hund's coupling parameters of the metals appear reasonable. Furthermore, the scalar approach and scaled $2\times2$ (atom-wise inversion) yield the same result across the board, in keeping with the conclusions of subsection~\ref{subsec:comparisons_with_conventional_approach}. The \edit{scaled $2\times2$} approach is \edit{marginally} less numerically stable, which is reflected by the marginally larger error estimates. \edit{Interestingly, however, we find that for the spin channel that matters to strong correlation (the spin-up channel for less-than-half filled sub-shells, and the spin-down channel for more-than-half filled sub-shells), the relevant $1\times1$ $U$ is very reasonable, and systematically lower in value than the conventional scalar $U$. This hints at a possible solution for first-principles DFT\,+\,$U$ calculations on systems in which the calculated scalar $U$ proves to be unphysically large, and the predominantly empty/full spin channel is already well described by the approximate functional.}

One particularly noteworthy result is the substantial spin-screening of the Hubbard parameters of [Co(H\textsubscript{2}O)\textsubscript{6}]\textsuperscript{3+} observed in averaged and non-averaged $1\times1$. This is the only complex in a low-spin ground state, so the up and down Kohn-Sham orbitals overlap perfectly and there is very efficient screening between spins. This system also exhibits one of the largest $J$ values. Similarly, the large $J$ values on the oxygen atoms may surprise at first (as Hund's physics is expected to play a very minor role here). This illustrates an important point: the absence of any magnetization does not imply the absence of magnetization-related error \edit{in the approximate functional}. Subsequent calculations demonstrate that applying this $J$ term, large as it is, does not result in the oxygen atoms acquiring magnetic moments.

Some works go one step further and calculate Hubbard parameters in a self-consistent fashion,\cite{Kulik2006a, Kulik2008a, Kulik2010a} with linear response being performed on DFT\,+\,$U$ ground states. While it remains to be seen what effect this additional step would have, it will \edit{likely be small here} because these systems do not undergo qualitative changes in electronic structure upon the application of $U$:\cite{Kulik2015b} in going from DFT to scalar DFT\,+\,$U$, the root-mean-square and maximum \edit{fractional} differences in the total $3d$ occupancies are \edit{6\% and 15\%} respectively. For the spin moment $\mu = n^\uparrow - n^\downarrow$ these are \edit{7\% and 14\%} respectively.

It is important to acknowledge that the authors of Ref.~\onlinecite{Zhao2016a} calculated $U$ for this set of \edit{molecules} (using scalar linear response). In comparison, their values are lower (by 1.4\,eV on average) and more species-dependent (a standard deviation of 1.2\,eV compared to 0.9\,eV for our set of values). \edit{In comparison with this work, Ref.~\onlinecite{Zhao2016a} (a) used ultra-soft pseudopotentials as opposed to norm-conserving ones; (b) performed all calculations on structures optimized in the $3+$ charge state; (c) employed $U$ self-consistency for some calculations; and (d) used of SCF linear response}. As the following section will demonstrate, details such as (a) and (b) can substantially affect Hubbard parameters.

\subsubsection{Comparison of schemes}
Table~\ref{tab:Udet_metal} illustrates the dangers of averaging across the two spin channels, as performed in averaged $1\times1$. For systems where both the spin-up and spin-down channels are partially occupied (see Fig.~\ref{fig:combined_U_J_and_occ}a) the responses are well-behaved, the Hubbard parameters are both sensible and similar, and averaging is unlikely to have any drastic effects. But for the heavier elements with filled spin-up channels, we are faced with the prospect of averaging two very different values, which in the most extreme cases lead to negative Hubbard parameters. Here, averaging the two values is likely to be an extremely poor approximation.

However, any Hubbard correction will not directly affect a fully-occupied channel, because the Hubbard energy correction term (Eq.~\ref{eqn:DFT+U_energy}) vanishes regardless of the magnitude of $U$. If it is imperative that the same correction must be applied to both channels, an argument could be made in favor of applying the $U^\downarrow$ value in place of an average. Of course, the Hubbard \emph{potential} does not vanish (Eq.~\ref{eqn:DFT+U_potential}) and fictional spin-up Kohn-Sham orbitals that overlap with the Hubbard projectors would be shifted by $U^\downarrow$. This inconsistency may have unforeseen effects\edit{, and an alternative may be to apply DFT\,+\,$U$ to partially-filled spin channels only.}

Table~\ref{tab:Udet_metal} also demonstrates the shortcomings of simple $2\times2$, the approximate atom-wise-inversion-based method. In the upper half of the table it yields reasonable values similar to those of scaled $2\times2$. But in the latter half (where dramatically different response in the spin-up and spin-down channels is expected) the approximation is a very poor one and the resulting parameters are unphysical. Scaled $2\times2$ encounters no such difficulties, justifying the use of the rescaling factors $\lambda_{U/J}$. This work will consider simple $2\times2$ no further.

\subsubsection{Dependence on simulation settings}
\begin{figure}[t]
\includegraphics[width=3.5in]{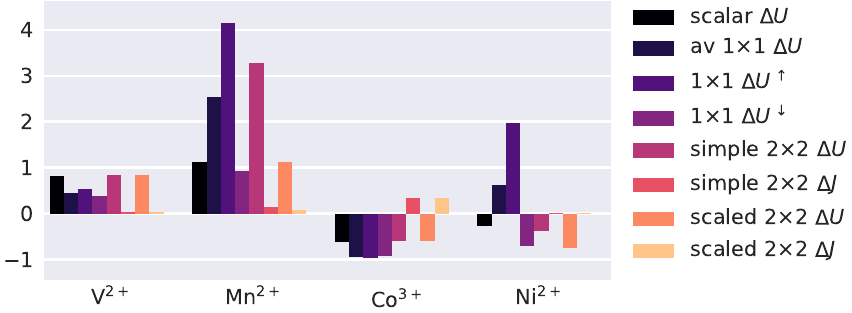}
\caption{\edit{The difference in Hubbard parameters for four hexahydrated transition metals, as calculated via the various linear response schemes and using two alternative simulation set-ups (eV).}}
\label{fig:DOR_vs_opium_bar_plot}
\end{figure}

The results of linear response calculations are sensitive to the precise settings of a calculation. Figure~\ref{fig:DOR_vs_opium_bar_plot} shows the difference in Hubbard parameters as obtained using two quite different simulation schemes. Both sets of calculations were performed on the same physical systems, but they differed in (a) the pseudopotentials used (Rappe \textit{vs.} in-house); (b) the electrostatic truncation scheme used (padded cell with a spherical cutoff\cite{Hine2011b} \textit{vs.} a Martyna-Tuckerman correction\cite{Martnya1999a}); and (c) the resolution of the fine grid used for calculating products of basis functions (a factor of two \textit{vs.} a factor of four finer than the standard grid). The majority of the Hubbard parameters match to within 1\,eV, except for those that relate to the response of a nearly-fully occupied subspace, where the response is \edit{extremely} changeable.

\subsubsection{A closed-shell system}
Linear response calculations were also performed on [Zn(H\textsubscript{2}O)\textsubscript{6}]\textsuperscript{2+}. Zn\textsuperscript{2+} is not strictly a transition metal, as its $3d$ shell is filled. Linear response calculations on closed shell systems tend to be troublesome,\cite{Hu2006a,Yu2014a} \edit{possibly due the response becoming non-linear.\cite{Huang2017a}}

\begin{table}[t!]
\caption{Values of $U$ and $J$ (eV) for the $3d$ subspace of Zn in hexahydrated zinc, calculated using the various linear response schemes and two alternative sets of Hubbard projectors (as defined by the net charge configuration of the Zn atom in a pseudoatomic solver).}
\label{tab:Udet_Zn_metal}
\footnotesize
\begin{tabular}{c c |@{\hspace{2ex}}
d @{$\pm$}
m{0.8cm} |@{\hspace{2ex}}
d @{$\pm$}
m{0.8cm}}
\hline
\multicolumn{2}{c |@{\hspace{2ex}}}{PAO charge} &
\multicolumn{2}{c |@{\hspace{2ex}}}{\hspace{-2ex}+0} &
\multicolumn{2}{c}{\hspace{-2ex}+2} \\
\hline
scalar                                 & $U$             & 10.05 & 0.03 & 34.77 & 0.01 \\
averaged $1\times1$                    & $U$             & 11.60 & 0.04 & 44.64 & 0.02 \\
\multirow{2}{*}{$1\times1$}            & $U^\uparrow$    & 11.67 & 0.06 & 44.65 & 0.03 \\
                                       & $U^\downarrow$  & 11.53 & 0.06 & 44.63 & 0.02 \\
\multirow{2}{*}{simple $2\times2$}     & $U$             & 10.08 & 0.03 & 34.79 & 0.02 \\
                                       & $J$             &  1.75 & 0.05 &  1.47 & 0.03 \\
\multirow{2}{*}{scaled $2\times2$}     & $U$             & 10.08 & 0.03 & 34.79 & 0.02 \\
                                       & $J$             &  1.75 & 0.05 &  1.47 & 0.03 \\
\hline
\end{tabular}

% 
% \begin{tabular}{m{1.4cm} |@{\hspace{2ex}}
% d @{$\pm$} 
% m{0.8cm} |@{\hspace{2ex}}
% d @{$\pm$}
% m{0.8cm} |@{\hspace{2ex}}
% d @{$\pm$}
% m{0.8cm}
% d @{$\pm$}
% m{0.8cm} |@{\hspace{2ex}}
% d @{$\pm$}
% m{0.8cm}
% d @{$\pm$}
% m{0.8cm} |@{\hspace{2ex}}
% d @{$\pm$}
% m{0.8cm}
% d @{$\pm$}
% m{0.8cm}
% d @{$\pm$}
% m{0.8cm}
% d @{$\pm$}
% m{0.8cm}}
% \hline
% \multicolumn{1}{c |@{\hspace{2ex}}}{PAO}
% &\multicolumn{2}{c |@{\hspace{2ex}}}{scalar}
% &\multicolumn{2}{c |@{\hspace{2ex}}}{averaged $1\times1$}
% &\multicolumn{4}{c |@{\hspace{2ex}}}{$1\times1$}
% &\multicolumn{4}{c |@{\hspace{2ex}}}{simple $2\times2$}
% &\multicolumn{4}{c }{scaled $2\times2$} \\
% \multicolumn{1}{c |@{\hspace{2ex}}}{charge}
% &\multicolumn{2}{c|@{\hspace{2ex}}}{$U$}
% &\multicolumn{2}{c|@{\hspace{2ex}}}{$U$}
% &\multicolumn{2}{c}{$U^\uparrow$}
% &\multicolumn{2}{c|@{\hspace{2ex}}}{$U^\downarrow$}
% &\multicolumn{2}{c}{$U$}
% &\multicolumn{2}{c|@{\hspace{2ex}}}{$J$}
% &\multicolumn{2}{c}{$U$}
% &\multicolumn{2}{c}{$J$} \\
% \hline
% \input{tab_Udet_Zn_PAOs_metal}
% \hline
% \end{tabular}
\end{table}

The results of our calculations are listed in Table~\ref{tab:Udet_Zn_metal}. These calculations were performed for two different definitions of the Hubbard projectors. In ONETEP these are defined using pseudoatomic orbitals (PAOs): that is, the DFT solutions of the isolated atom/ion with the pseudopotential.\cite{Sankey1989a,Artacho1999a,Ruiz-Serrano2012a} Table~\ref{tab:Udet_Zn_metal} lists the Hubbard parameters for when the pseudoatomic problem was solved with a total charge of 0 and +2\edit{, keeping the pseudopotential itself fixed}. The Hubbard projectors corresponding to the neutral pseudoatom are more diffuse than \edit{those for} the +2 case.

We find that $U$ is exceptionally large as given by both the scalar and spin-resolved linear response schemes, and with either definition of the Hubbard projectors. The dependence of the result on the Hubbard projectors is very striking, and is the most dramatic case that we have seen. But what is more remarkable is the robustness of these calculations (as shown by the small uncertainties). Crucially, this robustness is not due to the fact that some schemes avoid matrix inversion: the uncertainties are similar for schemes where matrix inversion is necessary ($2\times2$) and those where it is not ($1\times1$), \edit{and in no case did we observe evidence of non-linear response.} % Instead, it can only be attributed to our revised definition of $\chi_0$.

%%%%%%%%%%%%%%%%%%%%%%%%%%%%%%%%%%%%%%%%%%%%%%%%%%%%%%%%%%%%%%%%%%%%%%%%%%%%%%%
\subsection{\edit{Properties of MnO}}
\label{subsec:MnO_properties}
%%%%%%%%%%%%%%%%%%%%%%%%%%%%%%%%%%%%%%%%%%%%%%%%%%%%%%%%%%%%%%%%%%%%%%%%%%%%%%%
\begin{figure}[t!]
   \includegraphics[width=\columnwidth]{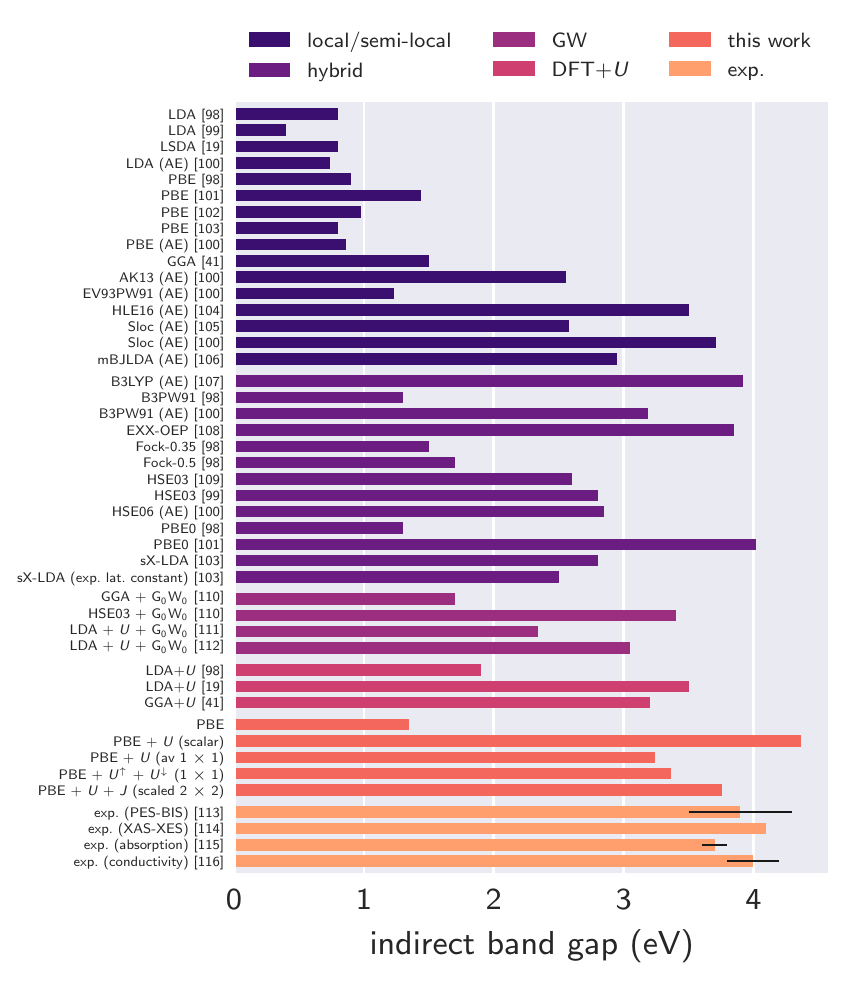}
   \caption{\edit{The indirect band gap of MnO, as calculated by various computational approaches, as well as experimental results (with error bars). All-electron calculations are denoted ``AE".}}
   \label{fig:MnO_bandgap}
\end{figure}

% List of citations in figure
\nocite{Tran2006a, Marsman2008, Anisimov1997a, Tran2017, Franchini2005a, Gopal2017a, Gillen2013a, Wang2006a, Verma2017, Finzel2017, Tran2009, Feng2004a, Engel2009a, Schron2010a, Rodl2009a, Jiang2010a, Kobayashi2008a, VanElp1991a, Kurmaev2008a, Iskenderov1969a, Drabkin1969a}

\begin{figure}[t!]
   \includegraphics[width=\columnwidth]{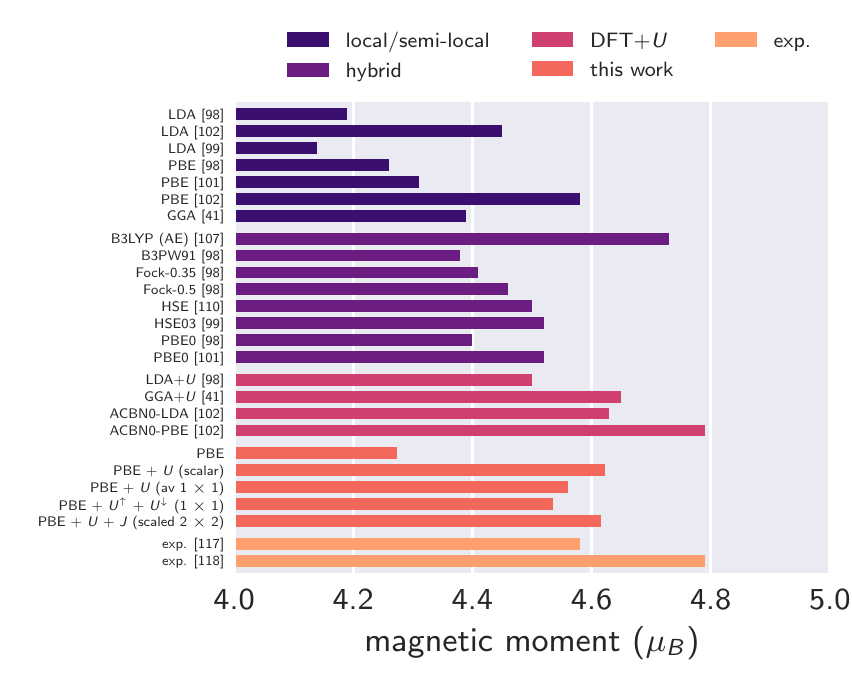}
   \caption{\edit{The magnetic moment of the manganese atoms in MnO, as calculated by various approaches.}}
   \label{fig:MnO_magmom}
\end{figure}

\edit{We calculated the band gap (Fig.~\ref{fig:MnO_bandgap}) and the local magnetic moment of Mn (Fig.~\ref{fig:MnO_magmom}) for bulk MnO using Hubbard and Hund's parameters obtained via our novel schemes (and listed in Tables~\ref{tab:Udet_metal} and \ref{tab:Udet_oxygen}). Semi-local functionals dramatically underestimate the band gap of MnO; the local/semi-local results presented in Fig.~\ref{fig:MnO_bandgap} underestimate it by $2.3$\,eV on average (with a standard deviation of $1.0$\,eV). They also underestimate the local magnetic moment (by $0.35 \pm 0.14$\,$\mu_B$). More sophisticated techniques have been applied with mixed success: hybrid, GW, and other DFT\,+\,$U$ studies underestimate the band gap by $1.3 \pm 1.0$, $1.3 \pm 0.7$, and $1.1 \pm 0.7$\,eV respectively. Our approaches compare very favourably, with the band gap agreeing with experiment, differing on average by $-0.2 \pm 0.4$\,eV. Scaled $2\times2$ in particular gives both band gap and magnetic moment in excellent agreement with experiment.}

\edit{It is worth mentioning that we found the predicted band gap to be highly sensitive to the choice of pseudopotential, with different pseudopotentials predicting anything from a metal to gaps as large as 2\,eV (for PBE). All-electron calculations yield a gap of 0.86\,eV.\cite{Tran2017} To obtain similar values with a pseudopotential, ensuring accurate $4s$ and $4p$ scattering proved to be key.}

% List of citations in figure
\nocite{Tran2006a, Gopal2017a, Marsman2008, Franchini2005a, Wang2006a, Feng2004a, Rodl2009a, Cheetham1983a, Fender1968a}

% \begin{figure}[t!]
%    \includegraphics[width=\columnwidth]{fig_MnO_ldos.pdf}
%    \caption{The indirect bandgap of MnO, as calculated by various computational approaches.}
%    \label{fig:MnO_bandgap}
% \end{figure}

\edit{Transition metal oxides are typically insulating for one of two reasons. Early $3d$ transition metal oxides (such as TiO and VO) are Mott-Hubbard insulators, with the band gap sitting between the lower and upper Hubbard bands. Late $3d$ transition metal oxides (such as CuO and NiO) are charge-transfer insulators, with band gaps formed between the oxygen $2p$ band and the upper metal $3d$ band, separated by the ligand-to-metal charge transfer energy.}

\begin{figure}[t!]
   \includegraphics[width=\columnwidth]{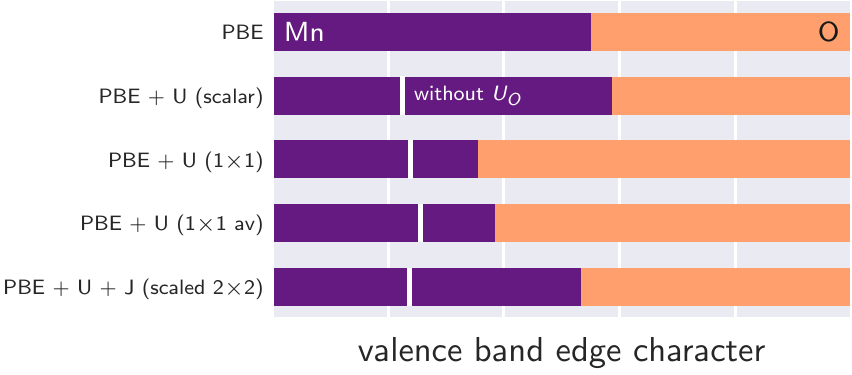}
   \caption{\edit{The valence band edge character of MnO, showing the fractional contribution of Mn (purple) and O (orange). PBE correctly predicts the valence band edge's mixed character, as do the different corrective schemes. This balance is due largely to the $U$ (and $J$ where relevant) terms applied to the oxygen $2p$ subspaces, which see the Mn fractions increase from unphysically low values (indicated in white).}}
   \label{fig:MnO_valence_band_edge_character}
\end{figure}

\begin{figure*}[t!]
   \includegraphics[width=\textwidth]{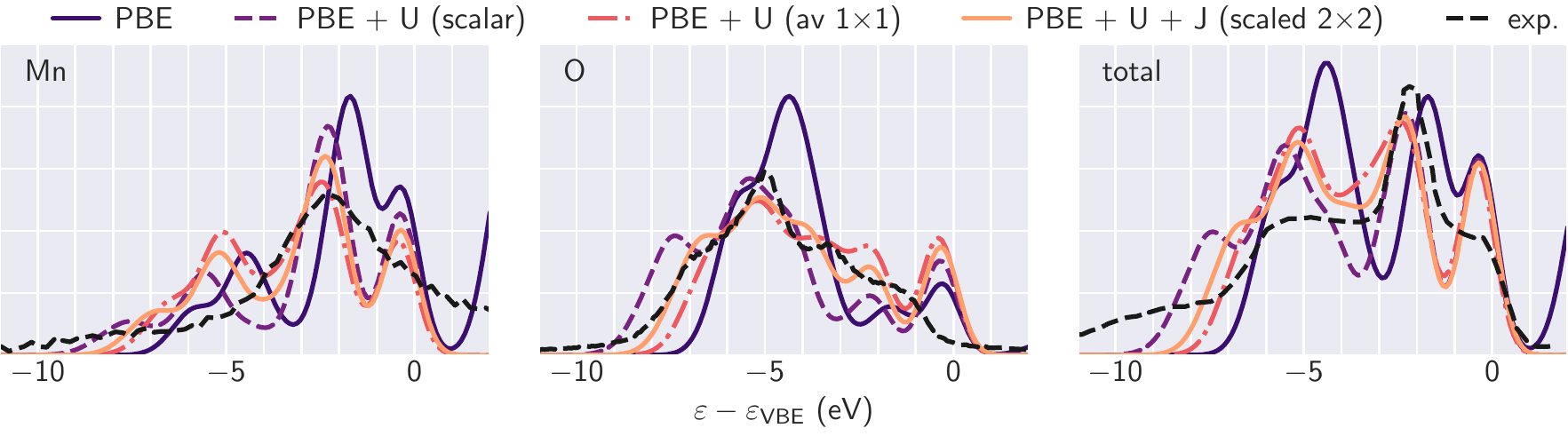}
   \caption{\edit{The local Mn, O, and total densities of states as obtained by the different schemes. The $1\times1$ result is similar to the averaged $1\times1$ result, and so has been excluded for simplicity. The energy scale is shown relative to the valence band edge energy $\varepsilon_\mathrm{VBE}$. Experimental results (XES and XPS) from Ref~\onlinecite{Kurmaev2008a} are included for comparison.}}
   \label{fig:MnO_all_ldos}
\end{figure*}

\edit{MnO sits near the boundary of these two regimes; the valence band edge is neither purely metal $3d$ or oxygen $2p$ in character.\cite{Fujimori1990a,VanElp1991a} As Fig.~\ref{fig:MnO_valence_band_edge_character} illustrates, this picture is captured by all schemes, with the valence band edge character sitting between 36 to 59\,$\%$ Mn. That said, if Hubbard corrections are applied to Mn but not O, the Mn character drops to below 26\% in all cases, incorrectly approaching the charge-transfer insulation. This demonstrates the importance of applying corrections to the oxygen orbitals. The valence band in its entirety is plotted in Fig.~\ref{fig:MnO_all_ldos}, and our methods exhibit marked improvement over PBE.}

% \edit{When we calculate the valence band edge character, ... text discussing Figure~\ref{fig:MnO_valence_band_edge_character}.}

%%%%%%%%%%%%%%%%%%%%%%%%%%%%%%%%%%%%%%%%%%%%%%%%%%%%%%%%%%%%%%%%%%%%%%%%%%%%%%%
\subsection{Structural properties \edit{of hexahydrated metal complexes}}
\label{subsec:hexX_structural}
%%%%%%%%%%%%%%%%%%%%%%%%%%%%%%%%%%%%%%%%%%%%%%%%%%%%%%%%%%%%%%%%%%%%%%%%%%%%%%%
% \subsubsection{Structural properties}
We will now examine how these various Hubbard corrections affect the resulting geometry of the hexahydrated metal systems. Hartree Fock,\cite{Akesson1992a} hybrid DFT,\cite{Kallies2001a,Yang2014a} and semi-local xc-functionals (such as PBE)\cite{Li1996a} already predict bond lengths consistent with experiment,\cite{Beattie1981a} without any need for Hubbard corrections. However, these corrections can dramatically affect structural properties; it would be undesirable for them to do so here.

\begin{figure*}[t!]
   \includegraphics[width=\linewidth]{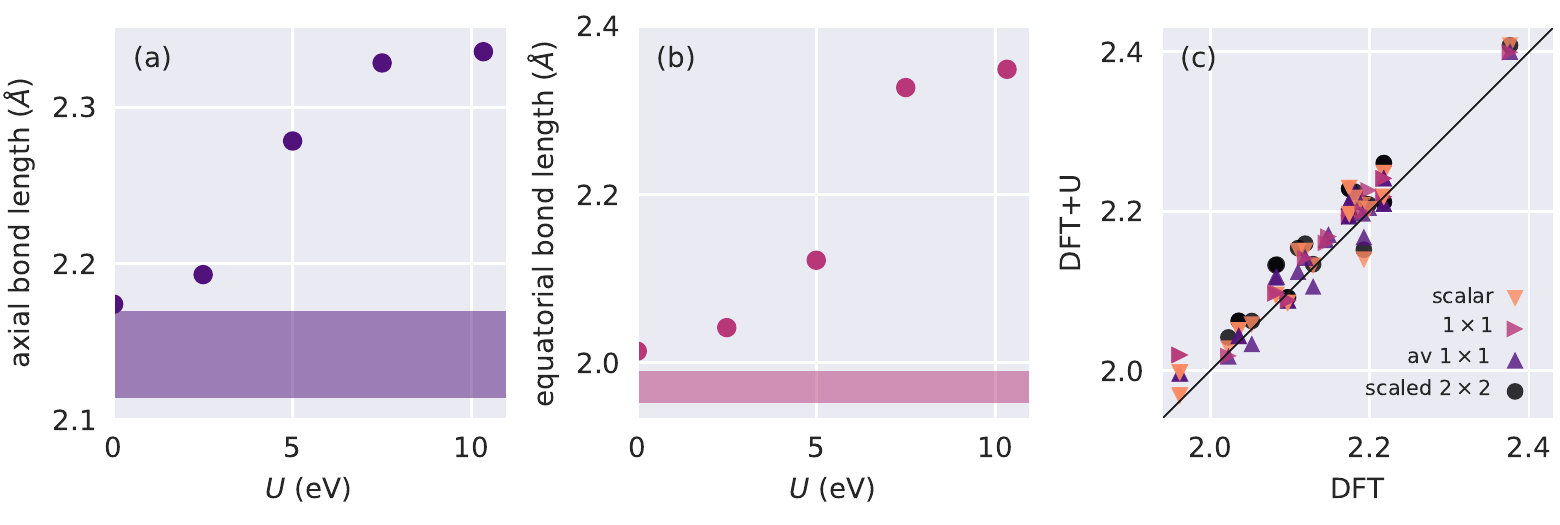}
   \caption{The mean (a) axial and (b) equatorial bond lengths of hexahydrated Mn\textsuperscript{3+} when optimized using DFT\,+\,$U$, for various values of $U^\text{Mn}$ and without adding a Hubbard correction to the oxygen atoms. The shaded regions indicate the range of values reported by other computational studies,\cite{Akesson1992a, Li1996a, Kallies2001a,Yang2014a} which are in line with experiment.\cite{Beattie1981a} (c) Metal-oxygen distances as given by DFT\,+\,$U$--optimized structures, \edit{now with a first-principles Hubbard $U$ correction to the oxygen $2p$ orbitals,} as compared to analogous PBE calculations. Each data-point corresponds to a distinct set of Hubbard parameters from Tables~\ref{tab:Udet_metal} and \ref{tab:Udet_oxygen} (that is, all different transition metal species and schemes for computing Hubbard parameters).}
   \label{fig:geometry_plots}
\end{figure*}

If only the $3d$ orbitals of the transition metal species are subjected to a Hubbard correction, and the structure optimized, metal-oxygen distances dramatically lengthen (Figs.~\ref{fig:geometry_plots}a and \ref{fig:geometry_plots}b). This is because any hybridization that existed between the metal $3d$ orbitals with lone pairs on the water ligands is weakened by the lowering of the energy of any filled $3d$ orbitals. Consequently, the individual species are stabilized and they drift apart. It is clear that this elongation is wholly unphysical, taking bond lengths well outside of the range of experimental values. This failure is not specific to this particular system or any procedure for computing $U$, but is a well-documented problem.\cite{Kulik2011a,Kulik2011b,Kulik2015b,Scherlis2007a}

There are a number of approaches for correcting this issue. One solution is DFT\,+\,$U$\,+\,$V$, which adds an inter-site interaction term to the DFT\,+\,U energy functional that may correctively favor O\,($2p$)--metal\,($3d$) bonding.\cite{Campo2010a} Alternatively, adaptive Hubbard projectors can mitigate the problem, as they will be more delocalized and responsive to the bonding environment.\cite{ORegan2010a} But perhaps the most pragmatic approach is to add Hubbard corrections to the $2p$ orbitals of the oxygen atoms.\cite{Nekrasov2000a,Cao2008a} This lowers their energies to levels comparable with the $3d$ orbitals, re-establishing the possibility of hybridization.

The success of the latter method is demonstrated in Fig.~\ref{fig:geometry_plots}c, where the addition of these corrections reduces any bond elongation to at most a five percent increase (and in many cases much less). The alignment is particularly remarkable given the range of different $U$ and $J$ values being used.

It is important to note that adding Hubbard terms to the oxygen atoms (a) alters hydrogen-oxygen-hydrogen angles by less than 2\%, (b) alters oxygen-hydrogen bond lengths by approximately 1\%, and (c) does not result in the oxygen atoms acquiring a magnetic moment (the largest observed was 0.014\,$\mu_B$ for DFT and 0.073\,$\mu_B$ for DFT\,+\,$U$\,+\,$J$).

% \begin{table}[t]
% \centering
% \caption{Geometry of hexahydrated transition metal systems as optimized using various approachs.}
% %\label{}
% \footnotesize
% \input{tab_hexX_bond_lengths_different_methods.tex}
% \end{table}

%%%%%%%%%%%%%%%%%%%%%%%%%%%%%%%%%%%%%%%%%%%%%%%%%%%%%%%%%%%%%%%%%%%%%%%%%%%%%%%
\subsection{Spectroscopic properties \edit{of the hexahydrated metal complexes}}
\label{subsec:hexX_spectroscopic}
%%%%%%%%%%%%%%%%%%%%%%%%%%%%%%%%%%%%%%%%%%%%%%%%%%%%%%%%%%%%%%%%%%%%%%%%%%%%%%%

\begin{table*}[t!]
\footnotesize
\caption{Spin flip energies (eV) for various hexahydrated transition metal systems. The quantum chemistry results are from Ref.~\onlinecite{Yang2014a}, and the experimental results are from Ref.~\onlinecite{Jorgensen1962a} (and the references therein).}
\begin{tabular}{c | *{9}{D{.}{.}{3.4}}}
\hline
\multicolumn{1}{c|}{\multirow{2}{*}{metal}} & \multicolumn{1}{c}{\multirow{2}{*}{DFT}} & \multicolumn{4}{c}{DFT\,+\,$U$\,(+$J$)} & \multicolumn{1}{c}{\multirow{2}{*}{CASSCF}} & \multicolumn{1}{c}{\multirow{2}{*}{CASPT2}} & \multicolumn{1}{c}{\multirow{2}{*}{MRCI}} & \multicolumn{1}{c}{\multirow{2}{*}{exp}}\\  
 &  & \multicolumn{1}{c}{scalar} & \multicolumn{1}{c}{av $1\times1$} & \multicolumn{1}{c}{$1\times1$} & \multicolumn{1}{c}{scaled $2\times2$} &  &  &  & \\  
\hline
V$^{2+}$ & 1.06 & 1.10 & 1.09 & 1.28 & 1.10 & 2.01 & 1.89 & 1.98 & 1.62\\  
Cr$^{3+}$ & 1.11 & 1.04 & 1.11 & 1.33 & 1.04 & 2.41 & 2.23 & 2.35 & 2.60\\  
Mn$^{2+}$ & 2.16 & 2.41 & 2.39 & 2.41 & 2.42 & 3.42 & 2.91 & 3.25 & 2.34\\  
Co$^{2+}$ & 1.60 & 1.85 & 1.85 & 1.86 & 1.85 & 1.96 & 1.95 & 1.76 & 1.98\\  
Ni$^{2+}$ & 1.23 & 1.44 & 1.48 & 1.50 & 1.44 & 2.30 & 2.03 & 2.23 & 1.91\\  
\hline
\end{tabular}

\label{tab:ddExcitations_spin_flips_unchanged_U}
\end{table*}

Hubbard corrections have significant bearing on spectroscopic properties (given that to first order, they open a gap between the filled and unfilled Hubbard projectors). This \edit{section} will focus on $d$-$d$ excitation energies, where a single electron transitions between two $3d$ orbitals. While these transitions are formally \edit{dipole-dipole} forbidden by the Laporte \edit{selection} rule, they are allowed via vibronic coupling.\cite{Harris1978a}

The first subset of such transitions are those which involve the flip of the electron's spin. These transitions additionally violate spin selection rules, but vibronic coupling again means that they are observable (albeit weakly). The transition energies are simply calculated as the difference in the total energy between two DFT\,(\,+\,$U$) calculations where the total spin differs by $\hbar$. This was done without updating $U$ (for a brief discussion regarding the updating of $U$ see Appendix~\ref{sec:hexX_Udet_spinflipped}). As this approach relies only on the accuracy of the total energy, DFT alone (without a Hubbard correction) might give reasonable results. This is indeed what we find (Table~\ref{tab:ddExcitations_spin_flips_unchanged_U}). The results are relatively insensitive to the choice of Hubbard parameters. \edit{Surprisingly, the scalar and scaled $2\times2$ approaches yield near-identical results, despite the fact that the two approaches differ by the value for $J$ and share the same value for $U$. A Hund's correction ought to have a significant bearing on spin-flip energies, providing further evidence that the precise functional form of the $+J$ functional needs revision.}

The other possible $d$-$d$ excitations involve the transition of a single electron without changing its spin.  These transitions are spin-allowed, and thus will exhibit intensities between those of fully allowed and spin-forbidden transitions. The transition energies are calculated as the difference in energy of the corresponding Kohn-Sham orbitals, and are listed in Table~\ref{tab:ddExcitations_spin_conserving}.

We find that DFT and DFT\,+\,$U$\,(+\,$J$) have mixed success reproducing these transition energies. This not surprising. The energy of such transitions is \edit{instead directly} related to the calculated \edit{Kohn-Sham} band gap and, as such, DFT (with its well-known underestimation of the band gap) will not give accurate results. Hubbard corrections tend to correctly enlarge \edit{Kohn-Sham} band gaps, but there is no reason \emph{a priori} why the final gap it produces ought to be accurate.\cite{Zhao2016a} Ongoing efforts are being made to construct generalized DFT\,+\,$U$ theories that satisfy Janak's/Koopman's theorem.\cite{Janak1978a,Koopmans1934a,Moynihan2016a} These transition energies will also be highly sensitive to static correlation, a failing of DFT associated with multi-\edit{reference} ground states. This failing remains unaddressed and may be an important factor in the overestimation of transition energies of Ti\textsuperscript{2+}, Fe\textsuperscript{2+}, and Co\textsuperscript{2+}.\cite{Cohen2008a,Cohen2012a} \edit{Adapting DFT\,+\,\emph{U}-like functionals to correct both self-interaction and static correlation error is an area of active research.\cite{Bajaj2017}} Furthermore, the excitation energies shown have been computed \edit{using} a very simplistic approach, neglecting vibronic and solvation effects (among others), which would likely result in significant shifts.\cite{Radon2016a}

\begin{table*}[t!]
\caption{\edit{Kohn-Sham} transition energies (eV) for spin-conserving $d$-$d$ excitations. \edit{In all cases, corrective terms were applied to both the metal $3d$ and oxygen $2p$ subspaces.}}
\footnotesize
\begin{tabular}{c c | *{9}{D{.}{.}{3.4}}}
\hline
\multicolumn{1}{c}{\multirow{2}{*}{metal}} & final & \multicolumn{1}{c}{\multirow{2}{*}{DFT}} & \multicolumn{4}{c}{DFT\,+\,$U$\,(+\,$J$)} & \multicolumn{1}{c}{\multirow{2}{*}{CASSCF}} & \multicolumn{1}{c}{\multirow{2}{*}{CASPT2}} & \multicolumn{1}{c}{\multirow{2}{*}{MRCI}} & \multicolumn{1}{c}{\multirow{2}{*}{exp}}\\  
 & symmetry &  & \multicolumn{1}{c}{scalar} & \multicolumn{1}{c}{av $1\times1$} & \multicolumn{1}{c}{$1\times1$} & \multicolumn{1}{c}{scaled $2\times2$} &  &  &  & \\  
\hline
\multicolumn{1}{c}{\multirow{4}{*}{Ti$^{3+}$}} & \textsuperscript{1}B\textsubscript{2g} & 0.27 & 3.16 & 1.48 & 1.63 & 2.79 & 0.00 & \multicolumn{1}{c}{} & \multicolumn{1}{c}{} & \multicolumn{1}{c}{}\\  
 & \textsuperscript{1}B\textsubscript{3g} & 0.28 & 3.25 & 1.51 & 1.66 & 2.86 & 0.00 & \multicolumn{1}{c}{} & \multicolumn{1}{c}{} & \multicolumn{1}{c}{}\\  
 & \textsuperscript{1}A\textsubscript{g} & 1.94 & 3.98 & 2.81 & 2.92 & 3.76 & 1.69 & 1.71 & 1.76 & 2.16\\  
 & \textsuperscript{1}A\textsubscript{g} & 2.38 & 4.60 & 3.33 & 3.44 & 4.34 & 1.70 & 1.72 & 1.77 & 2.52\\  
\hline
\multicolumn{1}{c}{\multirow{3}{*}{V$^{2+}$}} & \textsuperscript{3}B\textsubscript{1g} & 1.97 & 4.92 & 3.98 & 4.35 & 4.61 & 1.19 & 1.26 & 1.28 & 1.53\\  
 & \textsuperscript{3}B\textsubscript{2g} & 1.97 & 4.92 & 3.98 & 4.35 & 4.61 & 1.19 & 1.26 & 1.28 & 1.53\\  
 & \textsuperscript{3}B\textsubscript{3g} & 1.97 & 4.92 & 3.98 & 4.35 & 4.61 & 1.19 & 1.26 & 1.28 & 1.53\\  
\hline
\multicolumn{1}{c}{\multirow{3}{*}{Cr$^{3+}$}} & \textsuperscript{3}B\textsubscript{1g} & 2.24 & 3.98 & 3.15 & 3.17 & 3.56 & 1.69 & 1.77 & 1.79 & 2.16\\  
 & \textsuperscript{3}B\textsubscript{2g} & 2.24 & 3.98 & 3.15 & 3.17 & 3.56 & 1.69 & 1.77 & 1.79 & 2.16\\  
 & \textsuperscript{3}B\textsubscript{3g} & 2.24 & 3.98 & 3.15 & 3.17 & 3.56 & 1.69 & 1.77 & 1.79 & 2.16\\  
\hline
\multicolumn{1}{c}{\multirow{4}{*}{Cr$^{2+}$}} & \textsuperscript{4}A\textsubscript{g} & 0.38 & 2.06 & 1.60 & 1.80 & 1.83 & 0.62 & 0.69 & 0.64 & 1.17\\  
 & \textsuperscript{4}B\textsubscript{2g} & 1.37 & 3.28 & 2.77 & 2.99 & 3.04 & 1.18 & 1.27 & 1.19 & \multicolumn{1}{c}{}\\  
 & \textsuperscript{4}B\textsubscript{3g} & 1.53 & 3.44 & 2.93 & 3.15 & 3.22 & 1.23 & 1.30 & 1.23 & \multicolumn{1}{c}{}\\  
 & \textsuperscript{4}B\textsubscript{1g} & 1.95 & 3.88 & 3.36 & 3.58 & 3.61 & 1.34 & 1.44 & 1.36 & 1.75\\  
\hline
\multicolumn{1}{c}{\multirow{4}{*}{Mn$^{3+}$}} & \textsuperscript{4}A\textsubscript{g} & 0.21 & 1.28 & 0.68 & 0.62 & 0.97 & 0.69 & 0.77 & 0.72 & 1.11\\  
 & \textsuperscript{4}B\textsubscript{2g} & 0.97 & 5.08 & 3.63 & 1.86 & 4.87 & 1.72 & 1.96 & 1.78 & 2.53\\  
 & \textsuperscript{4}B\textsubscript{3g} & 2.64 & 5.28 & 3.99 & 3.41 & 4.95 & 1.76 & 1.99 & 1.82 & 2.53\\  
 & \textsuperscript{4}B\textsubscript{1g} & 3.00 & 5.72 & 4.38 & 3.88 & 5.42 & 1.91 & 2.21 & 2.00 & 2.53\\  
\hline
\multicolumn{1}{c}{\multirow{4}{*}{Fe$^{2+}$}} & \textsuperscript{4}B\textsubscript{2g} & 1.28 & 5.62 & 6.03 & 4.92 & 4.95 & 0.00 & \multicolumn{1}{c}{} & \multicolumn{1}{c}{} & \multicolumn{1}{c}{}\\  
 & \textsuperscript{4}B\textsubscript{3g} & 1.28 & 5.63 & 6.04 & 4.93 & 5.04 & 0.00 & \multicolumn{1}{c}{} & \multicolumn{1}{c}{} & \multicolumn{1}{c}{}\\  
 & \textsuperscript{4}A\textsubscript{g} & 1.88 & 5.73 & 6.07 & 5.07 & 5.24 & 0.75 & 0.80 & 0.83 & 1.29\\  
 & \textsuperscript{4}A\textsubscript{g} & 3.12 & 6.99 & 7.34 & 6.38 & 6.33 & 0.85 & 0.89 & 0.91 & 1.29\\  
\hline
\multicolumn{1}{c}{\multirow{6}{*}{Co$^{2+}$}} & \textsuperscript{3}B\textsubscript{2g} & 3.02 & 7.85 & 8.65 & 7.03 & 7.16 & 0.00 & \multicolumn{1}{c}{} & \multicolumn{1}{c}{} & \multicolumn{1}{c}{}\\  
 & \textsuperscript{3}B\textsubscript{3g} & 3.03 & 7.85 & 8.67 & 7.04 & 7.17 & 0.00 & \multicolumn{1}{c}{} & \multicolumn{1}{c}{} & \multicolumn{1}{c}{}\\  
 & \textsuperscript{3}B\textsubscript{2g} & 3.64 & 7.91 & 8.75 & 7.15 & 7.23 & 0.67 & 0.81 & 0.65 & 1.02\\  
 & \textsuperscript{3}B\textsubscript{3g} & 3.64 & 7.91 & 8.78 & 7.16 & 7.24 & 0.68 & 0.81 & 0.65 & 1.02\\  
 & \textsuperscript{3}B\textsubscript{3g} & 5.17 & 9.41 & 10.08 & 8.76 & 8.69 & 2.82 & 2.69 & 2.62 & 2.41\\  
 & \textsuperscript{3}B\textsubscript{2g} & 5.17 & 9.42 & 10.10 & 8.77 & 8.70 & 2.85 & 2.69 & 2.62 & 2.41\\  
\hline
\multicolumn{1}{c}{\multirow{6}{*}{Ni$^{2+}$}} & \textsuperscript{2}B\textsubscript{3g} & 4.30 & 8.74 & 13.39 & 7.72 & 7.91 & 0.75 & \multicolumn{1}{c}{} & \multicolumn{1}{c}{} & \multicolumn{1}{c}{}\\  
 & \textsuperscript{2}B\textsubscript{1g} & 4.30 & 8.74 & 13.39 & 7.72 & 7.91 & 0.76 & 0.89 & 0.85 & 1.05\\  
 & \textsuperscript{2}B\textsubscript{2g} & 4.30 & 8.74 & 13.39 & 7.72 & 7.91 & 0.76 & \multicolumn{1}{c}{} & \multicolumn{1}{c}{} & \multicolumn{1}{c}{}\\  
 & \textsuperscript{2}B\textsubscript{1g} & 4.30 & 8.74 & 13.39 & 7.72 & 7.91 & 1.31 & 1.48 & 1.45 & 1.67\\  
 & \textsuperscript{2}B\textsubscript{2g} & 4.31 & 8.74 & 13.39 & 7.72 & 7.91 & 1.31 & \multicolumn{1}{c}{} & \multicolumn{1}{c}{} & \multicolumn{1}{c}{}\\  
 & \textsuperscript{2}B\textsubscript{3g} & 4.31 & 8.74 & 13.39 & 7.72 & 7.91 & 1.31 & \multicolumn{1}{c}{} & \multicolumn{1}{c}{} & \multicolumn{1}{c}{}\\  
\hline
\multicolumn{1}{c}{\multirow{4}{*}{Cu$^{2+}$}} & \textsuperscript{1}A\textsubscript{g} & 1.79 & 5.06 & 0.49 & 4.56 & 4.47 & 0.51 & 0.61 & 0.52 & 1.17\\  
 & \textsuperscript{1}B\textsubscript{2g} & 2.31 & 6.19 & 1.12 & 5.49 & 5.33 & 0.84 & 1.08 & 0.85 & \multicolumn{1}{c}{}\\  
 & \textsuperscript{1}B\textsubscript{3g} & 2.82 & 6.68 & 1.81 & 5.91 & 5.82 & 0.89 & 1.12 & 0.89 & \multicolumn{1}{c}{}\\  
 & \textsuperscript{1}B\textsubscript{1g} & 3.34 & 6.85 & 1.99 & 5.95 & 5.88 & 0.97 & 1.23 & 0.99 & 1.56\\  
\hline
\end{tabular}

\label{tab:ddExcitations_spin_conserving}
\end{table*}

%%%%%%%%%%%%%%%%%%%%%%%%%%%%%%%%%%%%%%%%%%%%%%%%%%%%%%%%%%%%%%%%%%%%%%%%%%%%%%%
\section{\label{sec:conclusions} Conclusions}
%%%%%%%%%%%%%%%%%%%%%%%%%%%%%%%%%%%%%%%%%%%%%%%%%%%%%%%%%%%%%%%%%%%%%%%%%%%%%%%
We have generalized the minimum-tracking linear response formalism for calculating $U$ and $J$ to multiple sites and spins.\cite{Moynihan2017a} In this formalism, the non-interacting response $\chi_0$ is strictly a ground state property. Previously, it was not possible to calculate Hubbard parameters via linear response in large, spin-polarized systems such as metalloproteins.\cite{Cole2012a,Cole2016a} But because minimum-tracking is compatible with direct minimization (common to linear-scaling density functional theory packages such as ONETEP), linear response calculations on large and complex systems are now possible. 

\edit{Crucially, this} formalism allowed us to work with spin relatively easily. We demonstrated that the scalar linear response approach, whose use is widespread, yields a Hubbard $U$ that is unscreened by the opposite spin channel of the same site. We presented alternative approaches that account for this screening. Specifically, the opposite spin channel can be included in the bath, which is consistent with the effective decoupling of spins into separate subspaces implied by the standard DFT\,+\,$U$ functional (\emph{i.e.} the $1\times1$ schemes). This lowers the resulting $U$ values. Alternatively (but not equivalently), if inter-spin interactions require correction then a Hund's coupling parameter ought to be used in conjunction with an adjusted Hubbard parameter (scaled $2\times2$).

Applying these approaches to hexahydrated transition metals revealed \edit{significant} trends in the Hubbard parameters \edit{across the transition metals}. The linear response calculations were remarkably stable numerically, offering a possible route forward for closed-shell solids. \edit{That said, the best DFT\,+\,$U$ like model, and hence the uniquely-defined linear-response calculation scheme for that model, seems to be difficult to predict for a given system and underlying exchange-correlation functional.}

\edit{In the case of MnO, a canonical strongly correlated system, our novel approaches gave band gaps, magnetic moments, and valence band edge characters in excellent agreement with experiment, with a satisfyingly small variance compared to hybrid functionals and other methods. In the case of the hexahydrated transition metal complexes all approaches reproduced reasonable bond lengths but none reliably reproduced experimental $d$-$d$ excitation energies. The $1\times1$ approach gave the best results for spin-flip energies (a well-defined ground-state property), but even these were not in very good agreement with quantum-chemistry results.} Here, it appears that the electronic structure appears to be too complicated to be accurately described by the standard DFT\,+\,$U$ functional, especially while static correlation remains unaddressed. \edit{This is an area of ongoing research.\cite{Bajaj2017} The development of DFT\,+\,$U$ methodologies are reliant on evermore accurate quantum chemistry benchmarks (\emph{e.g.} Refs. \onlinecite{Song2018,Phung2018}).}

Applying Hubbard corrections to the oxygen $2p$ subspaces proved to be necessary to \edit{preserve the correct valence band edge character in MnO} and to reproduce bond lengths in hexahydrated transition metals.

By establishing a systematic approach for including/excluding screening by the opposite spin channel, these developments provide a route forward for performing DFT\,+\,$U$\,(\,+\,$J$) on spin-polarized systems in a robust and consistent manner.

\begin{acknowledgments}
E.\,B.\,L.\ acknowledges financial support from the Rutherford Foundation Trust and the EPSRC Centre for Doctoral Training in Computational Methods for Materials Science under grant EP/L015552/1. The calculations were funded by EPSRC Grant EP/J017639/1 and were performed using the Darwin Supercomputer of the University of Cambridge High Performance Computing Service (http://www.hpc.cam.ac.uk/), provided by Dell Inc.\ using Strategic Research Infrastructure Funding from the Higher Education Funding Council for England and funding from the Science and Technology Facilities Council. 

 The authors thank G.\ Moynihan, O.\ K.\ Orhan, and N.\ D.\ M.\ Hine for useful discussions.
\end{acknowledgments}

\appendix
%%%%%%%%%%%%%%%%%%%%%%%%%%%%%%%%%%%%%%%%%%%%%%%%%%%%%%%%%%%%%%%%%%%%%%%%%%%%%%%
\section{\label{sec:chi0_in_detail} Details of linear response theory}
%%%%%%%%%%%%%%%%%%%%%%%%%%%%%%%%%%%%%%%%%%%%%%%%%%%%%%%%%%%%%%%%%%%%%%%%%%%%%%%
\edit{We outline here the standard formalism for linear-response density-functional theory, following Refs.~\onlinecite{Gross1985,Baroni2001,Marques2004} and many others.}

Suppose for a given system we perturb the external potential by some small $\delta v_\text{ext}(\br)$. The resulting change in the density is given by
\begin{equation}
\delta n(\br) = \int d\br' \chi(\br, \br') \delta v_\text{ext}(\br')
\label{eqn:proof_chi}
\end{equation}
where $\chi(\br, \br')$ is the response function to this perturbation. For the same perturbation, we can choose to define a second response function $\chi_0(\br, \br')$ as
\begin{equation}
\delta n(\br) = \int d\br' \chi_0(\br, \br') \delta v_\mathrm{KS}(\br').
\label{eqn:proof_chi0}
\end{equation}
The Kohn-Sham potential is given as $v_\mathrm{KS}(\br) = v_\mathrm{Hxc}[n](\br) + v_{\text{ext}}(\br)$ --- that is, the sum of the Hartree and exchange-correlation potential, and the external potential (which includes the atomic potentials as well as the perturbing potential). It follows that $\delta v_\mathrm{KS}(\br) = \delta v_\mathrm{Hxc}[n](\br) + \delta v_{\text{ext}}(\br)$. By the Kohn-Sham construction, the change in the Hubbard-plus-xc-potential can be recast as
\begin{equation}
\delta v_\mathrm{Hxc}[n](\br') = \int d\br'' f[n_{GS}](\br',\br'') \delta n(\br''),
\label{eqn:proof_VHxc}
\end{equation}
where we have defined the Hartree plus exchange-correlation kernel as
\begin{equation}
f[n_{GS}](\br',\br'') = \left.\frac{\delta v_\mathrm{Hxc}(\br')}{\delta n(\br'')}\right|_{n=n_{GS}}.
\label{eqn:proof_fHxc}
\end{equation}
Combining \cref{eqn:proof_chi0,eqn:proof_chi,eqn:proof_VHxc,eqn:proof_fHxc} we can see that $\chi$, $\chi_0$, and $f$ are related via a Dyson-like equation for the Hartree plus exchange-correlation kernel:
\begin{widetext}
\begin{align}
\chi(\br, \br') =& \chi_0(\br, \br') + \int d\br'' \int d\br''' \chi_0(\br, \br''') f[n_{GS}](\br''',\br'') \chi(\br'', \br')
\end{align}
\end{widetext}
and we can identify $\chi_0(\br, \br') = \delta n(\br)/\delta v_\mathrm{KS}(\br')$ as the non-interacting response. \edit{For subspaces defined by projection operators $\hat{P}^J$, equation~\ref{eqn:variational_chi0} defines the projected non-interacting response, which is used in the minimum-tracking formalism for $U$ and in the present work.}

%%%%%%%%%%%%%%%%%%%%%%%%%%%%%%%%%%%%%%%%%%%%%%%%%%%%%%%%%%%%%%%%%%%%%%%%%%%%%%%
\section{\label{sec:the_lambda_approximation} Approximating $U$ and $J$ in the case of atom-wise inversion}
%%%%%%%%%%%%%%%%%%%%%%%%%%%%%%%%%%%%%%%%%%%%%%%%%%%%%%%%%%%%%%%%%%%%%%%%%%%%%%%
This appendix will cover the derivation of \cref{eqn:U_2x2a,eqn:U_2x2b,eqn:J_2x2a,eqn:J_2x2b}, with particular reference to the approximations involved.

Firstly, consider the Hubbard parameter. It was demonstrated (Eq.~\ref{eqn:U_as_an_exact_frac}) that it is given exactly by
% %
% \begin{equation}
% U =\frac{\delta^2 E_\mathrm{Hxc}}{\delta n^2} = \frac{\delta v_\mathrm{Hxc}}{\delta n} = \frac{1}{2}\frac{\delta v_\mathrm{Hxc}^\uparrow + \delta v_\mathrm{Hxc}^{\downarrow}}{\delta (n^{\uparrow} + n^{\downarrow})}
% \end{equation}
% %
% Using the result from linear response theory that $\delta v_\mathrm{Hxc}^\sigma = \sum_{\sigma'}f^{\sigma \sigma'} \delta n^{\sigma'}$ this can be reformulated as
% %
\begin{equation}
U = \frac{1}{2}\frac{f^{\uparrow\uparrow}dn^\uparrow +f^{\uparrow\downarrow}dn^\downarrow +f^{\downarrow\uparrow} dn^\uparrow +f^{\downarrow\downarrow}dn^\downarrow}{d(n^{\uparrow} + n^{\downarrow})}.
\label{eqn:U_fdn_form}
\end{equation}
We can interpret Eq.~\ref{eqn:U_fdn_form} as a statement that $U$ is given by a weighted average of the elements of $f^{\sigma \sigma'}$, where elements are weighted according to the extent to which the spin-up and -down densities would respond to a perturbation. In the case of spin-unpolarized systems, the two densities would respond equally ($dn^\uparrow = dn^\downarrow$) and Eq.~\ref{eqn:U_fdn_form} simplifies to
\begin{equation}
U = \frac{1}{2}(f^{\uparrow\uparrow} + f^{\uparrow\downarrow})
\end{equation}
(where we have also taken advantage of the symmetries $f^{\uparrow\uparrow} = f^{\downarrow\downarrow}$ and $f^{\uparrow\downarrow} = f^{\downarrow\uparrow}$). Such a straightforward simplification is not possible for spin-polarized systems. Instead, we must account for the possibility of different spin-up and -down density responses. To this end, we consider the ratio
\begin{equation}
 \frac{dn^\uparrow}{dn^\downarrow}
 = \frac{
   \sum_{\sigma} \chi^{\uparrow \sigma} dv^\sigma_\text{ext}
   }{
   \sum_{\sigma} \chi^{\downarrow \sigma} dv^\sigma_\text{ext}
   }.
\end{equation}
If we focus in particular on a perturbation of the form $dv^\uparrow_\text{ext} = dv^\downarrow_\text{ext}$ this simplifies to
\begin{equation}
 \frac{
   \sum_{\sigma} \chi^{\uparrow \sigma}
   }{
   \sum_{\sigma} \chi^{\downarrow \sigma}
   }
 = \lambda_U.
\end{equation}
Therefore, if we assert that in general $dn^\uparrow/dn^\downarrow$ can be approximated by $\lambda_U$ then Eq.~\ref{eqn:U_fdn_form} simplifies to 
\begin{align}
 U  =& \frac{1}{2}
   \frac{
      \lambda_U(
         f^{\uparrow\uparrow}
         + f^{\downarrow\uparrow}
         ) 
      + f^{\uparrow\downarrow}
      + f^{\downarrow\downarrow}
      }{
         \lambda_U + 1
      }.
\end{align}
This is Eq.~\ref{eqn:U_2x2b} of scaled $2\times2$. This approximation is reasonable but not rigorously justified, and is perhaps best interpreted \textit{post hoc}: for better or worse, scalar linear response makes this approximation (as demonstrated by the results of subsection~\ref{subsec:comparisons_with_conventional_approach}).

For the Hund's coupling parameter $J$ %= -d^2E_\mathrm{Hxc}/d\mu^2$
one can derive the analogous expression of Eq.~\ref{eqn:J_2x2a} in a very similar manner, except that the scaling factor $\lambda_J$ is constructed with reference to a perturbation of the form $dv^\uparrow_\text{ext} = -dv^\downarrow_\text{ext}$ (that is, one that will most directly affect magnetic moments).

Equations~\ref{eqn:U_2x2a} and \ref{eqn:J_2x2a} (simple $2\times2$) are more drastic approximations derived by assuming $\lambda_U = -\lambda_J= 1$. As the results of this paper demonstrate, these are poor approximations for spin-polarized systems.

%%%%%%%%%%%%%%%%%%%%%%%%%%%%%%%%%%%%%%%%%%%%%%%%%%%%%%%%%%%%%%%%%%%%%%%%%%%%%%%
\section{\label{sec:hexX_Udet_spinflipped} Linear response calculations for excited spin states}
%%%%%%%%%%%%%%%%%%%%%%%%%%%%%%%%%%%%%%%%%%%%%%%%%%%%%%%%%%%%%%%%%%%%%%%%%%%%%%%
\begin{table*}[t!]
\caption{Hubbard parameters calculated via linear response for systems where one electron's spin has been flipped from the ground spin state. The differences to the parameters obtained for the ground state (Table~\ref{tab:Udet_metal}) are listed in parentheses.} % \textcolor{red}{N.B. of these systems, Ref.~\onlinecite{Yang2014a} only provides spin-flip energies for V$^{2+}$, Cr$^{3+}$, Mn$^{2+}$, Fe$^{2+}$, Fe$^{3+}$, Co$^{2+}$, and Ni$^{2+}$ so try to focus on these systems}}
\footnotesize
\begin{tabular}{c |
D{.}{.}{4.3} 
D{.}{.}{3.5} |
D{.}{.}{4.3} 
D{.}{.}{3.5} |
D{.}{.}{4.3}
D{.}{.}{3.5}
D{.}{.}{4.3} 
D{.}{.}{3.5} |
D{.}{.}{4.3}
D{.}{.}{3.5}
D{.}{.}{4.3}
D{.}{.}{3.5}}
\hline
\multirow{2}{*}{\parbox{1.6cm}{metal}}
&\multicolumn{2}{c |}{scalar}
&\multicolumn{2}{c |}{averaged $1\times1$}
&\multicolumn{4}{c |}{$1\times1$}
&\multicolumn{4}{c }{scaled $2\times2$} \\
&\multicolumn{2}{c|}{$U$}
&\multicolumn{2}{c|}{$U$}
&\multicolumn{2}{c}{$U^\uparrow$}
&\multicolumn{2}{c|}{$U^\downarrow$}
&\multicolumn{2}{c}{$U$}
&\multicolumn{2}{c}{$J$} \\
\hline
\centering V$^{2+}$ & 3.99 & (-0.01) & 2.57 & (-0.21) & 2.72 & (-0.57) & 2.42 & (+0.14) & 3.84 & (-0.23) & 0.35 & (+0.01)\\
\centering Cr$^{3+}$ & 4.03 & (+0.13) & 1.69 & (-0.09) & 1.71 & (-0.15) & 1.68 & (-0.02) & 4.01 & (-0.03) & 0.41 & (+0.01)\\
\centering Cr$^{2+}$ & 3.08 & (-0.12) & 2.04 & (-0.35) & 2.12 & (-0.63) & 1.97 & (-0.07) & 3.08 & (-0.26) & 0.31 & (-0.02)\\
\centering Mn$^{3+}$ & 5.26 & (-0.14) & 1.64 & (-0.36) & 1.59 & (+0.08) & 1.69 & (-0.81) & 5.27 & (-0.59) & 0.50 & (+0.00)\\
\centering Mn$^{2+}$ & 4.33 & (-0.03) & 2.97 & (-1.08) & 3.19 & (-1.09) & 2.74 & (-1.08) & 4.56 & (-0.34) & 0.38 & (+0.01)\\
\centering Co$^{2+}$ & 5.11 & (+0.16) & 2.85 & (-3.34) & 2.86 & (-5.31) & 2.85 & (-1.37) & 5.12 & (-2.03) & 0.42 & (-0.06)\\
\centering Ni$^{2+}$ & 5.49 & (+0.23) & 3.32 & (-6.52) & 3.31 & (-12.10) & 3.32 & (-0.95) & 5.48 & (-6.87) & 0.90 & (+0.15)\\

\hline
\end{tabular}
\label{tab:Udet_spinflipped}
\end{table*}

\begin{table*}[t!]
\footnotesize
\caption{Spin flip energies (eV) for various hexahydrated transition metal systems with $U$ \edit{(and $J$ for $2\times2$)} updated following the flip. The quantum chemistry results are from Ref.~\onlinecite{Yang2014a}, and the experimental results are from Ref.~\onlinecite{Jorgensen1962a} (and the references therein). \edit{The surprisingly small (and even negative) spin-flip energies for $2\times2$ are strongly reminiscent of the findings of Millis and coworkers, who showed that the contemporary $+J$ functional wrongly disfavours ferromagnetism.\cite{Chen2016}}}
\begin{tabular}{c | *{9}{D{.}{.}{3.4}}}
\hline
\multicolumn{1}{c|}{\multirow{2}{*}{metal}} & \multicolumn{1}{c}{\multirow{2}{*}{DFT}} & \multicolumn{4}{c}{DFT\,+\,$U$\,(+$J$)} & \multicolumn{1}{c}{\multirow{2}{*}{CASSCF}} & \multicolumn{1}{c}{\multirow{2}{*}{CASPT2}} & \multicolumn{1}{c}{\multirow{2}{*}{MRCI}} & \multicolumn{1}{c}{\multirow{2}{*}{exp}}\\  
 &  & \multicolumn{1}{c}{scalar} & \multicolumn{1}{c}{av $1\times1$} & \multicolumn{1}{c}{$1\times1$} & \multicolumn{1}{c}{scaled $2\times2$} &  &  &  & \\  
\hline
V$^{2+}$ & 1.06 & 1.11 & 1.26 & 1.28 & 0.80 & 2.01 & 1.89 & 1.98 & 1.62\\  
Cr$^{3+}$ & 1.11 & 0.94 & 1.32 & 1.33 & -0.15 & 2.41 & 2.23 & 2.35 & 2.60\\  
Mn$^{2+}$ & 2.16 & 2.40 & 2.73 & 2.74 & 2.00 & 3.42 & 2.91 & 3.25 & 2.34\\  
Co$^{2+}$ & 1.60 & 1.83 & 1.44 & 1.62 & 0.72 & 1.96 & 1.95 & 1.76 & 1.98\\  
Ni$^{2+}$ & 1.23 & 1.81 & 1.03 & 1.50 & 0.41 & 2.30 & 2.03 & 2.23 & 1.91\\  
\hline
\end{tabular}

\label{tab:ddExcitations_spin_flips}
\end{table*}

The results of Table~\ref{tab:ddExcitations_spin_flips_unchanged_U} came from DFT\,+\,$U$ \edit{total energies for} each system in a high-spin and a lowered-spin state. Both of these calculations used the same Hubbard parameters, obtained via linear response calculations on the high-spin state.

Linear response calculations were also performed on the lowered-spin states (see Table~\ref{tab:Udet_spinflipped}). It was found that using updated Hubbard parameters worsened the resulting spin-flip energies (Table~\ref{tab:ddExcitations_spin_flips}), with some cases even predicting the wrong ground state.

%%%%%%%%%%%%%%%%%%%%%%%%%%%%%%%%%%%%%%%%%%%%%%%%%%%%%%%%%%%%%%%%%%%%%%%%%%%%%%%
\section{\label{sec:cRPA_comparison} A comparison with cRPA}
%%%%%%%%%%%%%%%%%%%%%%%%%%%%%%%%%%%%%%%%%%%%%%%%%%%%%%%%%%%%%%%%%%%%%%%%%%%%%%%
For the sake of comparison, it is instructive to study how constrained random-phase approximation (cRPA) methods account for the spin-screening of Hubbard parameters.\cite{Aryasetiawan2004a,Sakakibara2017a} In these approaches, the non-interacting response $\chi_0$ is partitioned into components corresponding to response within/between various subspaces. For instance, consider a system consisting of a single site with spin-up and -down channels. The component due to response solely within the spin-up subspace is given by the $(\uparrow, \uparrow)$th entry of $\chi_0$ --- that is,
\begin{equation}
(\chi_{0,\uparrow})^{\sigma \sigma'} \equiv
\begin{pmatrix}
{\chi_0}^{\uparrow \uparrow} & 0 \\
0 & 0
\end{pmatrix}.
\end{equation}
The non-interacting response due to all other contributions is
\begin{equation}
(\tilde \chi_{0,\uparrow})^{\sigma \sigma'} 
\equiv \chi_0 - \chi_{0,\uparrow}
=
\begin{pmatrix}
0 & {\chi_0}^{\uparrow \downarrow} \\
{\chi_0}^{\downarrow \uparrow} & {\chi_0}^{\downarrow \downarrow}
\end{pmatrix}.
\end{equation}
For such a non-interacting response $\tilde \chi_{0,\sigma}$ there is a corresponding Dyson equation
\begin{equation}
U^\sigma_{RPA} = \left[\left(f^{-1} - \tilde \chi_{0,\sigma}\right)^{-1}\right]^{\sigma \sigma}
\end{equation}
where $U^\sigma_{RPA}$ is now screened by everything save interactions within the spin-$\sigma$ subspace (as this screening is what $\tilde \chi_{0,\sigma}$ pertains to).

Screened interaction parameters $U^\sigma_{RPA}$ for hexahydrated metal systems are tabulated in Table~\ref{tab:Udet_cRPA_metal_diff}. In this work, it was shown that point-wise inversion (the averaged and non-averaged $1\times1$ schemes) yields an interaction screened by both the opposite spin channel on the same site and the remainder of the system, so we expect the results of Table~\ref{tab:Udet_cRPA_metal_diff} to resemble those of Table~\ref{tab:Udet_metal}. They are correlated, but the match is certainly not exact. This suggests that the RPA is not a good approximation for screening between unlike-spins, and that more sophisticated methods (such as that of Ref.~\onlinecite{Sasioglu2010a}) are required.
% 
% \begin{figure*}[t!]
% \includegraphics[width=7in]{fig_B_vs_cRPA_values_plot.pdf}
% \caption{Spin-screened Hubbard parameters $U^\sigma$ for the transition metal $3d$ subspaces, as calculated via point-wise inversion and cRPA. If the cRPA was exact, the two pairs of values would match.}
% \label{fig:B_vs_cRPA_values}
% \end{figure*}

\begin{table}[t!]
\centering
\caption{Spin-screened Hubbard parameters $U$ (eV) calculated using the cRPA approach. The differences with respect to the corresponding averaged and non-averaged $1\times1$ results of Table~\ref{tab:Udet_metal} are given in parentheses.}
\label{tab:Udet_cRPA_metal_diff}
\footnotesize
\begin{tabular}{c |
D{.}{.}{3.3}
D{.}{.}{3.4}
D{.}{.}{3.3}
D{.}{.}{3.4}
D{.}{.}{3.3}
D{.}{.}{3.4}}
\hline
\parbox{1.3cm}{metal}
&\multicolumn{2}{c}{average} 
&\multicolumn{2}{c}{$U^\uparrow$}
&\multicolumn{2}{c}{$U^\downarrow$} \\
\hline
\centering Ti$^{3+}$ & 0.80 & (-0.86) & 0.88 & (-0.97) & 0.71 & (-0.76)\\
\centering V$^{2+}$ & 2.22 & (-0.56) & 2.57 & (-0.72) & 1.88 & (-0.40)\\
\centering Cr$^{3+}$ & 1.16 & (-0.62) & 1.06 & (-0.80) & 1.26 & (-0.44)\\
\centering Cr$^{2+}$ & 2.07 & (-0.32) & 2.31 & (-0.44) & 1.83 & (-0.21)\\
\centering Mn$^{3+}$ & 1.17 & (-0.83) & 0.38 & (-1.13) & 1.95 & (-0.55)\\
\centering Mn$^{2+}$ & 3.47 & (-0.58) & 3.15 & (-1.13) & 3.78 & (-0.04)\\
\centering Co$^{3+}$ & 1.20 & (+0.01) & 1.20 & (+0.01) & 1.20 & (+0.01)\\
\centering Co$^{2+}$ & 5.19 & (-1.00) & 6.23 & (-1.94) & 4.14 & (-0.08)\\
\centering Ni$^{2+}$ & 8.36 & (-1.48) & 12.39 & (-3.02) & 4.32 & (+0.05)\\
\centering Cu$^{2+}$ & -3.53 & (-0.99) & -11.44 & (-2.33) & 4.37 & (+0.33)\\

\hline
\end{tabular}
\end{table}

%%%%%%%%%%%%%%%%%%%%%%%%%%%%%%%%%%%%%%%%%%%%%%%%%%%%%%%%%%%%%%%%%%%%%%%%%%%%%%%
% Bibliography
\end{document}